\documentclass[aps,jcp,preprint,superscriptaddress]{revtex4}

\usepackage[utf8]{inputenc}
\usepackage{physics}
\usepackage{amsmath, amssymb}
\usepackage{tabularx,booktabs,array,dcolumn}
\usepackage{setspace}
\usepackage{vmargin}

\usepackage{color}

\usepackage{graphicx,psfrag,subfigure}

\usepackage{float}
\usepackage[all]{xy}

\newcommand{\mx}[1]{\boldsymbol{#1}}

\newcommand{\cm}{cm$^{-1}$}

\def\lgmx{\mx{g}}

\def\m0{m^{(0)}}

\def\som{Supplementary Material}

\def\cthv{$C_\text{3v}$}
\def\ctwv{$C_\text{2v}$}
\def\td{$T_\text{d}$}
\def\chfar{CH$_4\cdot$Ar}
\def\aone{A$_1$}
\def\atwo{A$_2$}
\def\fone{F$_1$}
\def\ftwo{F$_2$}

\def\som{Supplementary Material}

\begin{document}

\title{%
Bound and unbound rovibrational states of the methane-argon dimer
}

\author{D\'avid Ferenc}

\author{Edit M\'atyus}
\email[matyus@chem.elte.hu]{}

\affiliation{Institute of Chemistry, E\"otv\"os Loránd University, Pázmány Péter sétány 1/A, Budapest, Hungary\\}
\date{September 27, 2018}

\begin{abstract}
\noindent %
Peculiarities of the intermolecular rovibrational quantum dynamics of the methane-argon complex are studied
using a new, \emph{ab initio} potential energy surface [Y. N. Kalugina, S. E. Lokshtanov, V. N. Cherepanov, and A. A. Vigasin, J. Chem. Phys. 144, 054304 (2016)],
variational rovibrational computations, and detailed symmetry considerations within the 
molecular symmetry group of this floppy complex
as well as within the point groups corresponding to the local minimum structures. 
The computed (ro)vibrational states up to and beyond the dissociation asymptote 
are characterized using two limiting models: the rigidly rotating molecule's model
and the coupled-rotor model of the rigidly rotating methane and an argon atom orbiting around it.
%
%
\end{abstract}

\maketitle

%
%
\section{Introduction}
\noindent Methane is an abundant molecule not only on Earth, but in the Universe \cite{Kalugina, Jager}. 
On Earth it has the third largest contribution to the greenhouse effect, after water vapor and carbon-dioxide, 
meanwhile on Titan, the moon of Saturn it is the most important greenhouse gas \cite{SAMUELSON}. 
Therefore, the interaction of methane with other simple molecules is of interest for astrochemistry 
and for atmospheric chemistry. There are several experimental and theoretical results for such systems including methane-water, CH$_4\cdot$H$_2$O \cite{ch4h2o2016,ch4h2o}, 
methane-fluoride, CH$_4\cdot$F$^-$, \cite{ch4f-},  and the mehane dimer, CH$_4\cdot$CH$_4$ \cite{ch4ch4}. 
\par 
The first measurements for the methane-argon gas mixture at several temperatures were reported by 
Dore and Filabozzi \cite{filabozzi}.
The infrared spectrum was reported first by McKellar \textit{et al.} in 1994 \cite{mckellar}, 
then Pak \textit{et al.} measured the high-resolution spectra of CH$_4\cdot$Ar and CH$_4 \cdot$Kr complexes in the 7 $\mu$m region. Later on, further high-resolution measurements and \textit{ab initio} computations were performed for the CH$_4\cdot$Ar system \cite{WANGLER, Jager}.   \par 

In spite of repeated attempts, the spectral lines within the bound region of the complex have remained elusive 
to experimentalists due to the very low induced dipole moment of the complex dominated by dispersion interactions 
(both isolated methane and isolated argon are apolar). This region is however well accessible to theoretical investigations
due to the low dimensionality (3D) of the intermolecular degrees of freedom and the excellent rigid-monomer approximation.
On the contrary, the experimentally well-accessible transitions to predissociative states of the complex are
extremely challenging for theory due to the high-dimensionality (12D) of the six-atomic methane-argon system. 
A rigorous full-dimensional rovibrational study has not been completed, and full-dimensional, 
\emph{ab initio} potential energy surface have not been developed yet. 
At the same time, the energy and lifetime of predissociative states carry ample
information on the fine details of the inter- and intra-molecular quantum dynamics taking place under 
the influence of weak (here: dispersion) interactions. 
Due to the very weakly bound nature of this complex and the high-order symmetry of the methane molecule, it has been anticipated that kinetic couplings play an important role in its intermolecular and predissociative dynamics. 

In the present work we take the opportunity to use a new, \emph{ab initio} atom-molecule interaction
potential energy surface of this dimer \cite{Kalugina} to study peculiarities of the rovibrational dynamics and related symmetry considerations.

%
%

\section{Computational details}
\begin{figure}
\begin{center}
  \includegraphics[scale=1.]{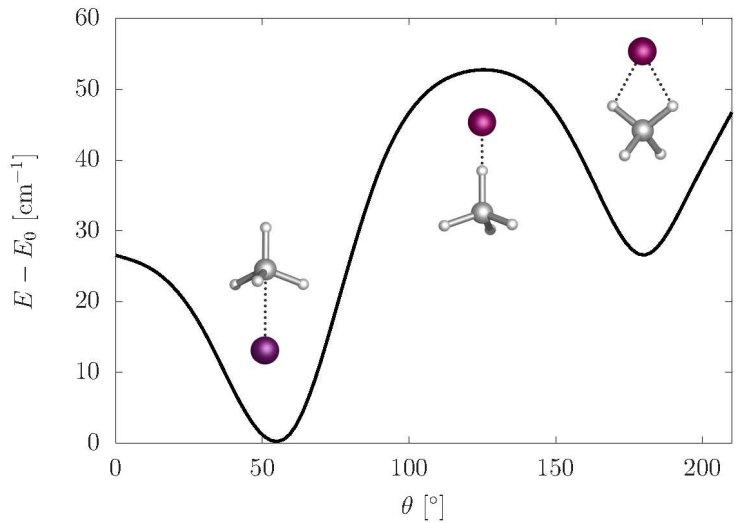}
\end{center}
\caption{Angular dependence of the potential energy surface of Ref.~\cite{Kalugina}
represented in terms of $(R,\theta,\phi)$ spherical polar coordinates. For each $\theta$ value in the plot the $R$ coordinate was chosen to minimize the interaction energy, while
$\phi$ was set to 0.0$^\circ$.
}
\label{figure01}
\end{figure}

\paragraph{Intermolecular potential energy surface}
We use the three--dimensional potential energy surface (PES) determined by Kalugina \emph{et al.} by \emph{ab initio} computations at the CCSD(T)/CBS level of theory, and by fitting an analytical functional form to the data points \cite{Kalugina}. 
The \emph{ab initio} computations were carried out with a regular tetrahedral methane
structure with $r_\text{CH} = 2.067 35$~bohr.
%
There are two non--equivalent minima on this PES: the argon atom is in a face configuration at the global minimum (GM) structure corresponding to an $R=6.95$~bohr 
carbon-argon distance, and in an edge configuration with a larger
$R=7.34$~bohr value at the secondary minimum (SM) structure. 
The saddle point connecting the two non-equivalent minima
is located at the vertex position of the argon atom (Fig.~\ref{figure01}). 
The dissociation energy ($D_e$) of the complex from the global minimum is 141.47~\cm\ and it is 
115.06~\cm\ from the secondary minimum. 

\paragraph{Rovibrational computations}
We carried out rovibrational computation on this intermolecular PES using the GENIUSH program \cite{geniush,geniush2}. 
GENIUSH stands for GENeral Internal-coordinate, USer-define Hamiltonians, and it
has been successfully used for the computation of bound and resonance states of floppy molecular systems,
including ArNO$^+$ \cite{ArNOp}, HeH$_2^+$ \cite{H2He}, NH$_3$ \cite{geniush,geniush2}, 
H$_5^+$ \cite{h5+}, CH$_5^+$ \cite{FaQuCs17}, CH$_4\cdot$H$_2$O \cite{ch4h2o,ch4h2o2016}, 
with various internal coordinate and frame definitions. An interesting feature, the existence of formally negative-energy rotational excitations 
\cite{ch5+,ch5+2,ch5+3,h5+,ch4h2o,ch4h2o2016} 
had been observed in a growing number of floppy systems and
led to the 
introduction of the notion \emph{astructural} molecules in Ref.~\cite{h5+}.
Ref.~\cite{h5+} also presented 
a one-dimensional torsional model to better highlight the phenomenon. 
Ref.~\cite{ch4h2o2016} reported an elaborate study on 
the quantum dynamical background of these
special features for the example of the methane-water dimer
using a quantum mechanical, coupled-rotor model of the subsystems. 

The general rovibrational Hamiltonian implemented in GENIUSH is
\begin{align}
\begin{split}
\hat{H}&=\frac{1}{2} \sum_{kl=1}^{D} \tilde{g}^{-1/4} \hat{p}^{\dagger}_k G_{kl} \tilde{g}^{1/2}  \hat{p}_l \tilde{g}^{-1/4} \\
&+ \frac{1}{2}\sum_k^D \sum_{\alpha}^3 \left( \hat{p}_k G_{k,D+\alpha} + G_{k,D+\alpha} \hat{p}_k   \right) \hat{J}_{\alpha}\\
&+ \frac{1}{2}\sum_\alpha^3 G_{D+\alpha,D+\alpha}\hat{J}_\alpha^2\\
&+\frac{1}{2} \sum_\alpha^3 \sum_{\beta>\alpha}^3 G_{D+\alpha,D+\beta}\left( \hat{J}_\alpha \hat{J}_\beta + \hat{J}_\beta \hat{J}_\alpha  \right)  + V ,
\label{eq:ham}
\end{split}
\end{align}
where $\hat{p}_k=-i\partial/\partial q_k$ is a differential operator corresponding to the $q_k$ internal coordinate, and $\hat{J}_\alpha$ ($\alpha=x,y,z$) labels the body-fixed angular momentum operators. The quantities $G_{kl}=(\lgmx^{-1})_{kl}$ and $\tilde{g}=\text{det}(\lgmx)$  are obtained from the $g_{kl}$ mass-weighted metric tensor, defined as
\begin{align}
g_{kl} &= \sum_{i=1}^N m_i \textbf{t}_{ik}^{\text{T}} \textbf{t}_{il}, && k,l \in \{1,2,...,D+3\}  \label{eq:gmx}\\
\textbf{t}_{ik} &= \frac{\partial \textbf{r}_i}{\partial q_k}, \label{eq:tvecvib}\\
\textbf{t}_{D+\alpha} &= \textbf{e}_{\alpha} \times \textbf{r}_i, && \label{eq:tvecrot} 
\end{align}
which are evaluated over a direct product grid in the program using the user-defined body-fixed Cartesian coordinates in terms of the internal coordinates. 

Geometrical constraints (in the present case for the methane molecule) are accounted for in an automated fashion by choosing $D$ being equal to the number of active vibrational dimensions ($D=3$ for the present case)
in Eq.~(\ref{eq:ham}). This choice is equivalent
to fixing the structural parameter (and a corresponding zero velocity) in the classical Lagrangian \cite{geniush}.
The Hamiltonian matrix is obtained as the matrix representation of Eq.~(\ref{eq:ham}) constructed with 
the discrete variable representation (DVR) for the vibrational degrees of freedom, and using the Wang functions for rotations. The kinetic energy operator is evaluated numerically over a direct product DVR grid. The eigenvalues and eigenfunctions of the Hamiltonian matrix are computed with an iterative Lanczos eigensolver. 

We define the body-fixed Cartesian coordinates for the methane-argon complex, as follows: we place the carbon atom in the origin, two hydrogen atoms are in the $y-z$ plane and the other two are in the $x-y$ plane with the same methane geometry as the PES \cite{Kalugina}. The position of the argon atom is defined by the $R,\cos\theta,$ and $\phi$ spherical polar coordinates, which are the
active vibrational coordinates (for the basis functions, detailed intervals and DVR parameters see Table~\ref{dvr}).
The coordinate definition is completed by shifting the system to the center of mass of the complex.
We used atomic masses \cite{NIST}, 
$m_\text{H} = 1.007\ 825\ 032\ 23$~u, 
$m_\text{D} = 2.014\ 101\ 778\ 12$~u,
$m_\text{T} = 3.016\ 049\ 277\ 9$~u,
$m_\text{C} = 12$~u, and 
$m_\text{Ar} = 39.962 383 1237$~u, throughout the computations.

\begin{table}
\caption{  
  Internal coordinate, basis function, number of DVR points ($N$), and grid intervals used in the computations.
}
\begin{center}
\begin{tabular}{@{}l l l l @{}}
\hline \hline
Coordinate & Basis function & $N$ & Grid interval\\
\hline 
$R$ [\r{A}] & Laguerre & 51 &  [2.65--15]  \\
$\cos \theta$ & Legendre &  81 & Unscaled \\
$\phi$ [$^\circ$] & Fourier &  21 &  [0--360) \\
\hline \hline
\end{tabular}
\end{center}
\label{dvr}
\end{table}

%
%

\section{Symmetry considerations and analysis tools \label{ch:symm}}

%
%

\noindent This section is dedicated to a general description of the symmetry analysis and 
symmetry properties of the limiting (rigid-rotor and coupled-rotor) models, 
which will be used to understand the numerical results in Section~\ref{ch:numres}.

\subsection[A]{Point groups and the  molecular symmetry group of the methane-argon dimer}
\begin{figure}[H]
\begin{center}
\includegraphics[width=.5\textwidth]{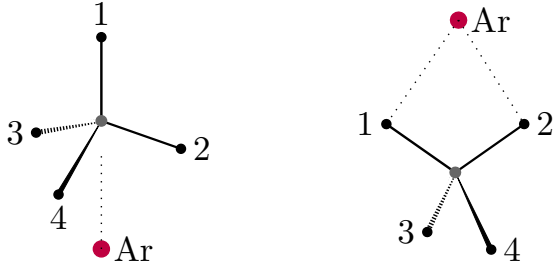}
\end{center}
\caption{The global minimum has $C_{3\text{v}}$ point-group symmetry, whereas the local minimum has $C_{2\text{v}}$ point-group symmetry (this particular numbering of the nuclei is used in the derivation in the text).}
\label{figure02}
\end{figure}

\noindent The PES of Ref.~\cite{Kalugina} features two non-equivalent minima. The global minimum has a \cthv\ point-group (PG) symmetry, while the secondary minimum has \ctwv\ PG symmetry (Figs.~\ref{figure01} and \ref{figure02}). With the low barriers and the almost freely rotating methane unit in the dimer, there is an interchange among the four GM and the six SM structures that gives rise to a rich splitting pattern. 

The methane-argon dimer belongs to the \td(M) molecular symmetry (MS) group and all symmetry species of the MS group are physically existing \cite{bunker-jensen}. 

\subsection[B]{General considerations about the tunneling splitting of vibrational states}
\noindent The splitting of vibrational states,
classified by PG irreducible representation (irrep) labels, 
will be derived by starting out from a simple picture.
Let us imagine that the system exhibits only small ($\varepsilon$) amplitude vibrations about the
minimum structures (of one of the PG symmetries). This system is almost ``frozen'' at the near proximity of these equilibrium structures, and thus its molecular symmetry group
is the permutation-inversion group isomorphic to the PG (with a selected numbering of the nuclei). Hence, these $\varepsilon$-vibrations can be characterized by irreps of the PG, \cthv (PG) or \ctwv (PG),
but they can also be characterized by the \cthv (M) and \ctwv (M) \emph{permutation-inversion} (PI) groups. 
These PI groups share elements with the MS group, which will be used later.

For a general derivation of the vibrational splittings due to the large-amplitude motions, we proceed as follows. First, we construct the symmetry-adapted basis functions for the irreps of the PI group (isomorphic with the PG) of the minimum structure. Then, we consider the transformation properties of these functions under the symmetry operations of the MS group, determine the characters for each class of the MS group, and reduce the representation.
We carry out the derivation for a general pair of MS and PI ($\sim$PG) groups, labelled
with $G$ and $g$, respectively. At the end of the derivation, we calculate the splittings for the case of methane-argon.
\begin{enumerate}
\item
  Let us consider a function $f_0$ localized at an arbitrarily selected version of the molecule. 
\item
  Map the $f_0$ basis function onto all feasible versions, $f_i = \hat{O}_i f_0$, by using all $\hat{O}_i$ $i \in \{1,2, \cdots , N\}$ MS group operations ($N=|G|$ is the order of $G$).  
\item
  It is always possible to choose $f_0$ in such way that $\braket{f_i}{f_j}=\delta_{ij}$ ($i,j=1,\ldots,N)$). \par 
\item
  Construct a linear combination of the $f_i$ $(i=1,\ldots,N)$ functions
  which transform according to the $\mathit{\Gamma}$ irrep of $g$:
\begin{align}
  \phi_i^{(\mathit{\Gamma})} = \frac{1}{n} \sum_{k=1}^n c_k^{(\mathit{\Gamma})} \hat{P}_k f_i,
  \quad i=1,\ldots,N
\label{basis}
\end{align}
where $\hat{P}_k \in g$ and $n=|g|$ is the order of $g$. 
\item
In order to determine the characters in $G$,
the trace of the matrix representation of $\hat{O}_R\in G$ is calculated
corresponding to the $\lbrace \phi_i^{(\mathit{\Gamma})}; i=1,\ldots,N \rbrace$ 
set of functions:
\begin{align}
\begin{split}
  \chi^{(\mathit{\Gamma})} (\hat{O}_R) 
  &= 
  \sum_{i=1}^N 
  \bra{\phi_i^{(\mathit{\Gamma})}}\hat{O}_R \ket{\phi_i^{(\mathit{\Gamma}})} \\
  &= 
  \frac{1}{n^2} 
  \sum_{i=1}^N \sum_{k=1}^n 
  \sum_{l=1}^n 
     c_k^{(\mathit{\Gamma})} 
     c_l^{(\mathit{\Gamma})*} \bra{\hat{P}_k f_i}\hat{O}_R \ket{\hat{P}_l f_i}
\end{split}
\label{char}
\end{align}
\item
In order to evaluate the trace in Eq.~(\ref{char}), we first identify the non-vanishing integrals by reordering the operators as
\begin{align}
 \bra{\hat{P}_k f_i}\hat{O}_R \ket{\hat{P}_l f_i} = \bra{f_i} \hat{P}_k^{-1}\hat{O}_R \hat{P}_l\ket{ f_i}.
 \label{integral}
\end{align}
Since $g\subset G$ and due to the closure property of a group,
$\hat{P}_k^{-1}\hat{O}_R \hat{P}_l\in G$. Since the set of (localized) $f_i$ functions 
was created by mapping $f_0$ with all possible operations of $G$, 
and the functions are orthogonal, 
a non-vanishing integral is obtained only if the product of the three operators,
$\hat{P}_k^{-1}$, $\hat{O}_R$, and $\hat{P}_l$ equal 
the identity:
\begin{align}
 \hat{P}_k^{-1}\hat{O}_R \hat{P}_l = \hat{I}\\
 \hat{O}_R = \hat{P}_k \hat{P}_l^{-1}.
\end{align}
\item
  The integral is non-vanishing for a selected $\hat{O}_R\in G$, if 
  $\exists \hat{P}_r \in g$: $\hat{P}_r=\hat{O}_R$:
  \begin{align}
    \sum_{k=1}^n\sum_{l=1}^n \bra{f_i} \hat{P}_k^{-1}\hat{O}_R \hat{P}_l\ket{ f_i}  = n.
  \end{align}
  Then, if  $\mathit{\Gamma}$ is the totally symmetric irrep (assume that it is labelled with $\text{A}_1$) of $g$, all $c_k^{(\text{A}_1)}$ characters equal 1, and we obtain:
  \begin{align}
   \chi^{\mathit{\Gamma}} (\hat{O}_R) &= \frac{1}{n^2} \sum_{i=1}^N \sum_{k=1}^n \sum_{l=1}^n c_k^{\text{A}_1} c_l^{\text{A}_1}  \bra{\hat{P}_k f_i}\hat{O}_R \ket{\hat{P}_l f_i}\\
   &= \begin{cases}
       \frac{1}{n^2} N n = \frac{N}{n}, \text{\ if} \ \exists \ \hat{P}_r \in g: \hat{P}_r  = \hat{O}_R \\
       0, \text{\ otherwise}
      \end{cases}
  \end{align}
  For a different labeling of the atoms the $\hat{O}_R$ operators in the same class have different characters, thus we average over the $ \chi (\hat{O}_R)$ values within a class of the group. (Equivalently, we could have considered all possible labeling of the molecule at the beginning.) Let us denote the class of $\hat{O}_R$ as $\mathcal{R}$ and $K_\mathcal{R}$ is the number of elements in this class of $G$. Then, we obtain
  \begin{align}
    \chi^{\mathit{\Gamma}} (\mathcal{R}) 
    &= \frac{1}{K_\mathcal{R}}  \sum_{\hat{O}_R \in \mathcal{R}} \chi(\hat{O}_R)  \\
    &= \begin{cases}
        \frac{k_r}{K_\mathcal{R}} \frac{N}{n},  \text{\ if} \ \exists \ 
        \hat{P}_r \in g: \hat{P}_r  = \hat{O}_R \in \mathcal{R}
        \\
        0, \text{\ otherwise,}
  	 \end{cases}
  \end{align}
  and $k_r$ is the number of operations in the $\mathcal{R}$ class of $G$ 
  for which 
  $\exists \ \hat{P}_r\in g: \hat{P}_r  = \hat{O}_R$. 
  It can be seen that this $k_r$ number of operations equals the number of elements 
  in the class of $g$ which contains $\hat{P}_r$. 
\item
  In general for a $\mathit{\Gamma}$ irrep of $g$, 
  the characters for the $\mathcal{R}$ class of $G$ are 
  \begin{equation}
    \chi^{(\mathit{\Gamma})}(\mathcal{R})
    = \begin{cases}
      c_r^{(\mathit{\Gamma})}\frac{k_r}{K_\mathcal{R}} \frac{N}{n}, 
      \text{\ if} \ \exists \ \hat{P}_r \in g: \hat{P}_r  = \hat{O}_R \in\mathcal{R}
      \\
      0, \text{\ otherwise}.
    \end{cases}
  \label{finalequation}                        
  \end{equation}
\end{enumerate}

\subsection[C]{Mathematical structure of the tunneling splittings in methane-argon}
First, we calculate the tunneling splitting of the vibrational states for the global minimum
structure.
The order of $T_{\text{d}}(\text{M})$ is $N=24$. The global minimum (GM) has \cthv \ PG symmetry, and
the order of $C_{3\text{v}}$ (PG or PI) is $n=6$. The equivalent operations and the orders of classes 
in the two group are collected in Table \ref{eqvoper}. 

\begin{table}
\caption{Equivalent operations in the $T_{\textnormal{d}}$(M) and $C_{3\text{v}} (\text{M})$ symmetry groups for the labeling shown in Fig.~\ref{figure02}.\label{tab:equivgroup1}}
\begin{center}
\begin{tabular}{c c c c c c}
\hline \hline
$T_{\textnormal{d}}$(M)&$E$ &(123) &(14)(23)&[(1423)]*&[(23)]*\\
$K_R$                  &1   &  8   &   3    &    6    &   6   \\
$C_{3\text{v}}$(M)     &$E$ & (123)&        &         &[(23)]*\\
$k_r$                  & 1   &  2   &        &         &   3  \\
\hline \hline
\end{tabular}
\end{center}
\label{eqvoper}
\end{table}

The \td(M) irreps for the vibrational splittings of states corresponding to 
the global minimum and the $C_{3\text{v}}$ PG (and its isomorphic MS) group are obtained from 
Eq.~(\ref{finalequation}) and Table~\ref{tab:equivgroup1}:
\begin{align}
\begin{split}
\mathit{\Gamma}_{\text{A}_1}^{C_{3\text{v}}} &=  \textnormal{A}_1 \oplus \textnormal{F}_2  \\
\mathit{\Gamma}_{\text{A}_2}^{C_{3\text{v}}} &=  \textnormal{A}_2 \oplus  \textnormal{F}_1  \\
\mathit{\Gamma}_{\text{E}}^{C_{3\text{v}}} &= \textnormal{E} \oplus \textnormal{F}_1 \oplus \textnormal{F}_2 , 
\end{split}
\label{c3vsplit}
\end{align}
which defines the mathematical structure of the splitting pattern for the vibrational states
assignable to the GM.

\begin{table}[h]
\caption{Equivalent operations in the $T_{\textnormal{d}}$(M) and $C_{2\text{v}} (\text{M})$ symmetry groups for the labeling shown in Fig.~\ref{figure02}.
\label{tab:equivgroup2}}
\begin{center}
\begin{tabular}{c c c c c c}
\hline \hline
$T_{\textnormal{d}}$(M)&$E$ &(123) &(12)(34)&[(1234)]*&[(12)]*\\
$K_R$                  &1   &  8   &   3    &    6    &   6   \\
$C_{3\text{v}}$(M)     &$E$ &      &(12)(34)&         &[(12)]*\\
$k_r$                  & 1  &      &   1    &         &   1  \\
\hline \hline
\end{tabular}
\end{center}
\label{eqvoper2}
\end{table}

We proceed similarly for the secondary minimum, use Eq.~(\ref{finalequation})
and Table~\ref{tab:equivgroup2} for the \ctwv\ group (of order $n=4$), 
and obtain the splitting pattern for the vibrational states
assignable to the SM as:
\begin{align}
\begin{split}
\mathit{\Gamma}_{\text{A}_1}^{C_{2\text{v}}} &= \text{A}_1 \oplus \text{E} \oplus \text{F}_2 \\
\mathit{\Gamma}_{\text{A}_2}^{C_{2\text{v}}} &= \text{A}_2 \oplus \text{E} \oplus \text{F}_1 \\
\mathit{\Gamma}_{\text{B}_1}^{C_{2\text{v}}} &= \text{F}_1 \oplus \text{F}_2 \\
\mathit{\Gamma}_{\text{B}_2}^{C_{2\text{v}}} &= \text{F}_1 \oplus \text{F}_2 .
\end{split}
\label{c2vsplit}
\end{align}

A difference in the mathematical structure of the GM and SM splittings is that the totally symmetric vibrations (such as the zero-point vibration) of the SM contain a doubly-degenerate species.

%
%

\subsection[C]{Coupled-rotor decomposition}
\noindent 
Molecular systems which can be separated into two rigidly rotating distinct parts, can be characterized by a model which includes only the coupling of the subsystems' angular momenta (and neglects any PES interactions). This scheme was introduced by Sarka \emph{et. al} in relation with the general, numerical kinetic-energy operator approach of GENIUSH and is called the coupled-rotor decomposition (CRD) \cite{ch4h2o}.

\begin{figure}
 \begin{center}
    \includegraphics[width=0.7\textwidth]{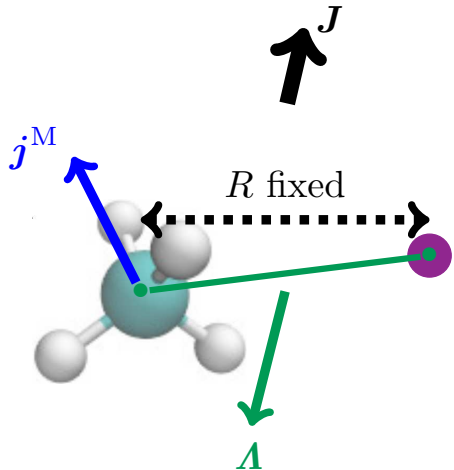}
 \end{center}
\caption{Coupling of the subsystems' angular momenta used to define the coupled-rotor model in CH$_4\cdot$Ar.}
\label{figure03}
\end{figure}

The coupled-rotor (CR) ansatz for methane-argon 
is constructed from the rotational wave functions of methane, $\phi_{km}^{j^{\text{M}}}$, and of the effective diatomic rotor, $Y_{\mathit{\Lambda}}^m$, coupled to a total rotational angular momentum state, $\ket{JM}$. Hence, the CR functions are defined as
\begin{align}
 [j^{\text{M}},\mathit{\Lambda}]_{JM}
 = \sum_{\mu}^{j^{\text{M}}}  \sum_{(M-\mu)}^{\mathit{\Lambda}}  \braket{j^\text{M},\mu,\mathit{\Lambda},(M-\mu)}{JM} \phi_\mu^{j^{\text{M}}} Y_{\mathit{\Lambda}}^{M-\mu},
 \label{crfunction}
\end{align}
where $\braket{j^\text{M},\mu,\mathit{\Lambda},(M-\mu)}{JM}$ are the Clebsch--Gordan coefficients.
Based on this definition, the symmetry properties of the rotational functions,
and the equivalent (monomer) rotations of the MS group operators,  
closed analytic expression can be derived for the symmetry labels of the CR functions. The characters and irrep decomposition in \td(M) 
are given in Tables~\ref{CRj0} and \ref{CRj1}
for the $J=0$ and $J=1$ CR functions of methane-argon.

\begin{table}
\caption{Characters and irrep decomposition of the $J=0$ rotational functions of the methane-argon dimer in the $T_{\textnormal{d}}$(M) molecular symmetry group. $n_\text{cl}$ is the number of operations in the class.}
\begin{center}
\begin{tabular}{c |c c c c c | c}
\hline \hline
& $E$ &  (123) & (14)(23) & [(1423)]* & [(23)]*&\\
$n_{\textnormal{cl}}$   &1&8&3&6&6& \\
\hline
$\mathit{\Gamma}$ &&&&&&Irreps \\
\hline
$[0,0]_{00}$& 1& 1& 1&1&1& A$_1$\\
$[1,1]_{00}$& 3& 0 &1&$-1$&1& F$_2$\\
$[2,2]_{00}$& 5& $-1$& 1&$-1$&1&E $\oplus$ F$_2$\\
$[3,3]_{00}$& 7&  1&$-1$& 1& 1&A$_1$ $\oplus$ F$_1$ $\oplus$ F$_2$\\
$[4,4]_{00}$& 9& 0 &1 &1 &1 & A$_1$ $\oplus$ E $\oplus$ F$_1$ $\oplus$ F$_2$ \\
$[5,5]_{00}$& 11& $-1$& $-1$& $-1$& 1 & E $\oplus$ F$_1$ $\oplus$ F$_2$ \\
$[6,6]_{00}$& 13& 1 &1 &$-1$ &1 & A$_1$ $\oplus$ A$_2$ $\oplus$ E $\oplus$ F$_1$ $\oplus$ 2 F$_2$\\ 
\hline \hline
\end{tabular}
\end{center}
\label{CRj0}
\end{table}

\begin{table}
\caption{Characters and irrep decomposition of the $J=1$ rotational functions of the methane-argon dimer in the $T_{\textnormal{d}}$(M) molecular symmetry group.}
\begin{center}
\begin{tabular}{c |c c c c c | c}
\hline \hline
& $E$ &  (123) & (14)(23) & [(1423)]* & [(23)]*&\\
$n_{\textnormal{cl}}$   &1&8&3&6&6& \\
\hline
$\mathit{\Gamma}$ &&&&&&Irreps \\
\hline
$[0,1]_{10}$&1  &1  &1   &$-1$ &$-1$  & A$_2$ \\
$[1,0]_{10}$&3  &0  &$-1$  &1  &$-1$  & F$_1$ \\
$[1,1]_{10}$&3  &0  &$-1$  &$-1$ & 1  & F$_2$\\
$[1,2]_{10}$&3  &0  &$-1$  &1  &$-1$  & F$_1$\\
$[2,1]_{10}$&5  &$-1$ &1   &1  &$-1$  & E $\oplus$ F$_1$\\
$[2,2]_{10}$&5  & $-1$&   1& $-1$& 1  & E $\oplus$ F$_2$\\
$[2,3]_{10}$&5  &$-1$  &1  &1  &$-1$  & E $\oplus$ F$_1$\\
$[3,2]_{10}$&7  &1  &$-1$  &$-1$  &$-1$  & A$_2$ $\oplus$ F$_1$ $\oplus$ F$_2$\\
$[3,3]_{10}$&7  & 1 & $-1$ & 1  & 1  &  A$_1$ $\oplus$ F$_1$ $\oplus$ F$_2$\\
$[3,4]_{10}$&7  &1  &$-1$  &$-1$  &$-1$  & A$_2$ $\oplus$ F$_1$ $\oplus$ F$_2$\\
$[4,3]_{10}$&9  &0  &1  &$-1$  &$-1$  &A$_2$ $\oplus$ E $\oplus$ F$_1$ $\oplus$ F$_2$\\ 
$[4,4]_{10}$&9  & 0 & 1 & 1  & 1  & A$_1$ $\oplus$ E $\oplus$ F$_1$ $\oplus$ F$_2$\\ 
$[4,5]_{10}$&9  &0  &1  &$-1$  &$-1$  & A$_2$ $\oplus$ E $\oplus$ F$_1$ $\oplus$ F$_2$\\ 
$[5,4]_{10}$&11  &$-1$  &$-1$  &1  &$-1$  & E $\oplus$ 2 F$_1$ $\oplus$ F$_2$\\ 
$[5,5]_{10}$&11  &$-1$  &$-1$  &$-1$  &1  & E $\oplus$ F$_1$ $\oplus$  2 F$_2$\\ 
\hline \hline
\end{tabular}
\end{center}
\label{CRj1}
\end{table}

In order to measure the similarity of the CR model and the ``exact'' intermolecular wave functions, we compute CRD overlaps defined as \cite{ch4h2o}
\begin{equation}
\text{CRD}^J_{nm}= \sum_{r=1}^{N_R} \abs{  \sum_{k=-J}^{J} \sum_{i=1}^{N_{\Omega}} \tilde{\Psi}^J_{nk}(\rho_r,\omega_i) \tilde{\varphi}^J_{mk}(\omega_i)}^2 .
\label{CRDcoeff}
\end{equation}
where $\tilde{\Psi}^J_{nk}(\rho_r,\omega_i)$ is the exact variational (ro)vibrational wave function in DVR, $\rho$ and $\omega$ denote the radial and (the collection of) angular coordinates respectively, and $\tilde{\varphi}^J_{mk}(\omega_i)$ are the CR functions depending only on the angular coordinates. By identifying the dominant values in the CRD matrix the assignment of the (ro)vibrational states is straightforward. 

In order to be able to numerically evaluate this integral in a ``black-box fashion'' 
of the GENIUSH protocol \cite{geniush}, we compute the CR functions 
in a reduced-dimensionality GENIUSH computation with 
the atom-molecule distance fixed at its equilibrium value, the PES 
switched off, and the same angular grid representation as for the ``full'' 
problem (Table~\ref{dvr}). The \td (M) irrep labels are assigned to the (ro)vibrational states in an almost automated fashion by inspecting the numerical CRD matrix and using the symmetry properties of the CR functions (see Tables \ref{CRj0}--\ref{CRj1}). 

%
%
\subsection[D]{Rigid-rotor decomposition}

In the traditional picture of molecular rotations and vibrations, molecular rotations \cite{bunker-jensen, wilson} are superimposed on the (higher-energy) molecular vibrations. The rigid-rotor decomposition (RRD) scheme was introduced \cite{RRD} to assign parent vibrational states to rovibrational wave functions. The RRD assignment is based on the overlap of a direct product basis of pure vibrational functions, $\psi^{J=0}_n$ and the $\phi^{\text{RR},J}$ rigid-rotor functions with the full rovibrational wave function, $\psi^{J>0}_n$ (computed with a selected body-fixed frame):
\begin{align}
\ket{\psi^{J=0}_n}  \otimes \ket{J,\phi^\text{RR}_l}  = \ket{\psi^{J=0}_n \cdot \phi^{\text{RR},J}_l} \label{RRDbasis}\\
S_{nl,m} = \braket{\psi^{J=0}_n \cdot \phi^{\text{RR},J}_l}{\psi^{J>0}_m}. \label{RRDcoef}
\end{align}
By computing the $S_{nl,m}$ overlap values, the dominant contribution of a vibrational state to a rovibrational state can be determined. 

In the present work, besides the analysis of the numerical RRD values, we would like to better understand the regularities in the RRD matrix introduced by symmetry relations. First of all, we determine the symmetry assignment of the $\phi^{\text{RR},J}$ functions in the \td (M) MS group. The equivalent rotations ($\alpha_0,\beta_0,\gamma_0$) for each class in the \td(M) are (123): $\left( 0,0,\frac{2\pi}{3} \right)$; (14)(23): $\left(\frac{\pi}{3},\pi,0 \right)$; [(1423)]*: $\left(\frac{\pi}{6},\frac{\pi}{2},-\frac{\pi}{6} \right)$; [(23)]*: $\left( 0,\pi , 0   \right)$.
The characters of this reducible representation are the traces of Wigner $D$ matrices.
\begin{equation}
\chi^j = \sum_{m=-j}^j D^{j}_{m,m}(\alpha_0,\beta_0,\gamma_0)
\end{equation}
The irrep decompositions for several $J$ quantum numbers are collected in Table \ref{methane_rrsym}.

\begin{table}[h]
\caption{Characters and irrep decomposition of the $\phi^{\text{RR},J}$ rigid-rotor functions in the $T_{\textnormal{d}}$(M) MS group.}
\begin{center}
\begin{tabular}{c |c c c c c | c}
\hline \hline
   & $E$ &  (123) & (14)(23) & [(1423)]* & [(23)]*&\\
$n_{\textnormal{cl}}$   &1&8&3&6&6& \\
\hline
$\mathit{J}$ &&&&&&Irreps \\
\hline
0&1& 1& 1& 1& 1&      A$_1$ \\
1&3& 0& $-1$& 1& $-1$&     F$_1$ \\
2&5& $-1$& 1& $-1$& 1&     E $\oplus$ F$_2$ \\
3&7& 1& $-1$& $-1$& $-1$&    A$_2$ $\oplus$ F$_1$ $\oplus$ F$_2$ \\
4&9& 0& 1& 1& 1&       A$_1$ $\oplus$ E $\oplus$ F$_1$ $\oplus$ F$_2$ \\
\hline \hline
\end{tabular}
\end{center}
\label{methane_rrsym}
\end{table}

Since the symmetry of the RRD basis functions, Eq.~(\ref{RRDbasis}), is determined
by the direct product of the $\mathit{\Gamma}(\psi^{J=0})$ and $\mathit{\Gamma}(\phi^{\text{RR},J>0})$ representations, the overlap in 
Eq.~(\ref{RRDcoef}) can only be non-zero if the direct product contains the irrep of the rovibrational state. The irrep decomposition in \td (M) of the direct product of the $\psi^{J=0}_n$ vibrational (A$_1$, A$_2$, E, F$_1$, F$_2$) and $\phi^{\text{RR},J>0}$,
which is F$_1$ for $J=1$:
\begin{align}
\begin{split}
\text{A}_1 \otimes \text{F}_1 &= \text{F}_1  \\
\text{A}_2 \otimes \text{F}_1 &= \text{F}_2  \\
\text{E}   \otimes \text{F}_1 &= \text{F}_1 \oplus \text{F}_2  \\
\text{F}_1 \otimes \text{F}_1 &=  \text{A}_1 \oplus \text{E} \oplus \text{F}_1 \oplus \text{F}_2  \\
\text{F}_2 \otimes \text{F}_1 &= \text{A}_2\oplus \text{E}\oplus\text{F}_1 \oplus\text{F}_2.
\label{rrdsym}
\end{split}
\end{align}
This result tells us, for example, that 
a rovibrational state of A$_2$ symmetry can only originate from an F$_2$ vibrational state.

%
%

\section{Numerical results\label{ch:numres}}

%
%

\subsection[A]{Vibrational states}
\begin{figure} 
\includegraphics[width=\textwidth]{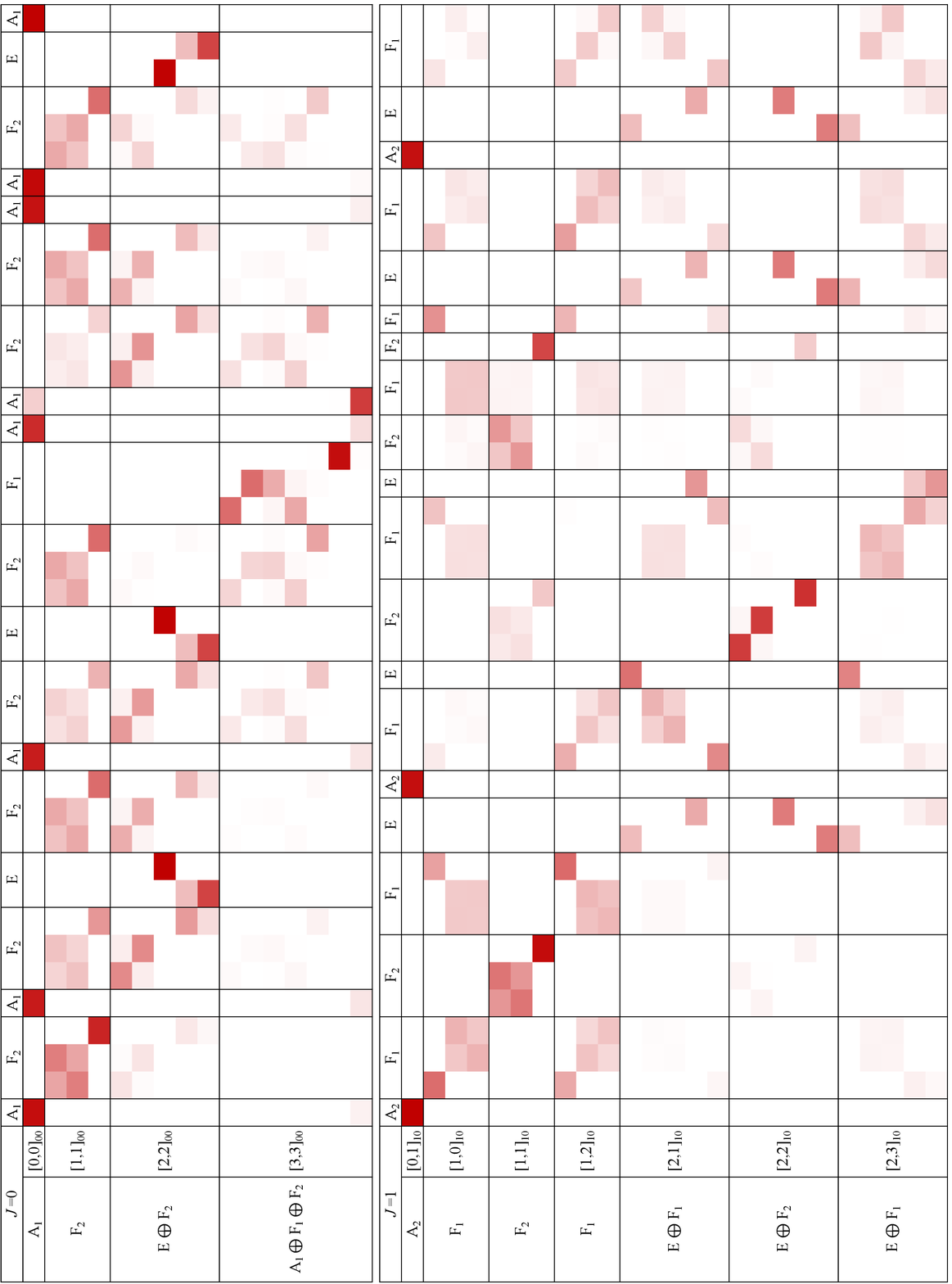}
\caption{CRD overlap coefficients, Eq. (\ref{CRDcoeff}), computed for the vibrational ($J=0$)
and rovibrational ($J=1$) states of CH$_4\cdot$Ar (a darker color corresponds to a larger value of the overlap). The rows correspond to the coupled-rotor functions $\varphi^J_n$, which are used to assign the full intermolecular vibrational states, $\Psi^J_n$, listed in the columns.}
\label{figure04}
\end{figure}

The depth of the PES valley of the global (secondary) minimum measured from the dissociation asymptote is $-D_e(\text{GM})= -141.47 $~\cm\ ($-D_e(\text{SM})= -115.06 $~\cm). 
By considering also the zero-point vibration (ZPV) of the GM, the lowest-energy state accessible to the system has an energy $-D_0 = -D_e + \text{ZPVE}=-88.60$~\cm, measured from the dissociation asymptote. 
The PES bounds 20 vibrational states (if we count degenerate states only once), which have been assigned using the limiting models described in Sec.~III. The computed energy levels, 
together with their CRD and \td (M) irrep assignments, are collected in Table~\ref{resoultsJ0} (see also Figure~\ref{figure04}).

\begin{figure}
  \begin{center}
   \includegraphics[width=1\textwidth]{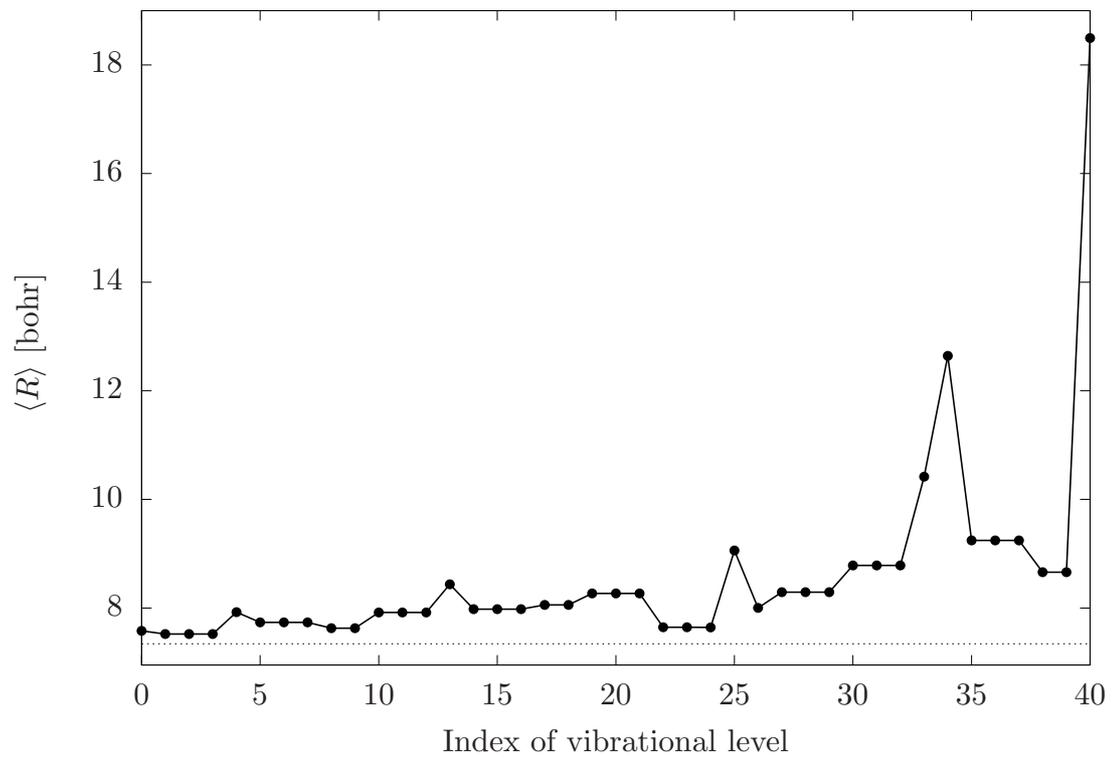}
  \end{center}
   \caption{%
     Expectation value of the carbon-argon distance, $R$, for the bound vibrational states of CH$_4\cdot$Ar. 
     \label{figure05}
   }
\end{figure}

First of all, by looking at the computed and assigned list of vibrational states, we have attempted to identify complete splitting patterns, which (would) 
correspond to the traditional picture of molecular vibrations split up
by feasible interchanges between the permutation-inversion versions.
Having in mind the mathematical structure of the splittings
for GM and SM vibrations, Eqs.~(\ref{c3vsplit}) and (\ref{c2vsplit}), 
we have inspected the expectation value of the atom-molecule separation over the bound vibrational states (Fig.~\ref{figure05}).
Due to the structural differences between GM and SM 
($R_\text{eq}(\text{GM})=6.95$~bohr and $R_\text{eq}(\text{SM})= 7.34 \ a_0$, respectively), an increased atom-molecule separation can be an indication of a secondary-minimum character.
Alternatively, an increased $\langle R\rangle$ value can also indicate vibrational excitation along the atom-molecule distance. 
Stretching vibrational excitations and overtones can be recognized by nodes
in the wave function along the $R$ coordinate. Figure~6 collects cuts of \aone-symmetry wave functions which highlight their nodal structure along $R$ (the \som\ contains similar cuts for the bound vibrational states). 

In addition, we repeated the vibrational computation also for the two symmetrically substituted isotopologues, CD$_4\cdot$Ar and CT$_4\cdot$Ar (Fig.~\ref{figure07}).
According to chemical-physical intuition, we expected that
the tunneling splittings should get smaller by
increasing the mass of the nuclei, which then could help the identification
of the vibrational splittings.

Based on all these considerations and the numerical results, 
the zero-point vibration of the global minimum, ZPV(GM), could be  
assigned to J0.1 (0.00~\cm, A$_1$) and J0.2--4 (9.01~\cm, F$_2$).
The next six states, J0.5 (28.74~\cm, \aone), J0.6--8 (31.24~\cm, F$_2$),
and J0.9--10 (32.08~\cm, E) could be assigned to the zero-point vibration
of the secondary minimum, ZPV(SM), but due to a node along the $R$ coordinate 
it would be morea appropriate to assign 
J0.5 (28.74~\cm, \aone) and J0.6--8 (31.24~\cm, F$_2$)
to  the intermolecular vibrational fundamental of the global minimum, Stre(GM).
In general, the overlap of the energy range of the ZPV(SM) and Stre(GM) (and other vibrations), and the appearance of the same symmetry species in the different splittings, Eqs.~(\ref{c3vsplit}) and (\ref{c2vsplit}), an unambiguous identification of vibrational states (in the traditional sense, split up by feasible exchanges) is hardly possible.

\begin{figure}[H]
\includegraphics[width=\textwidth]{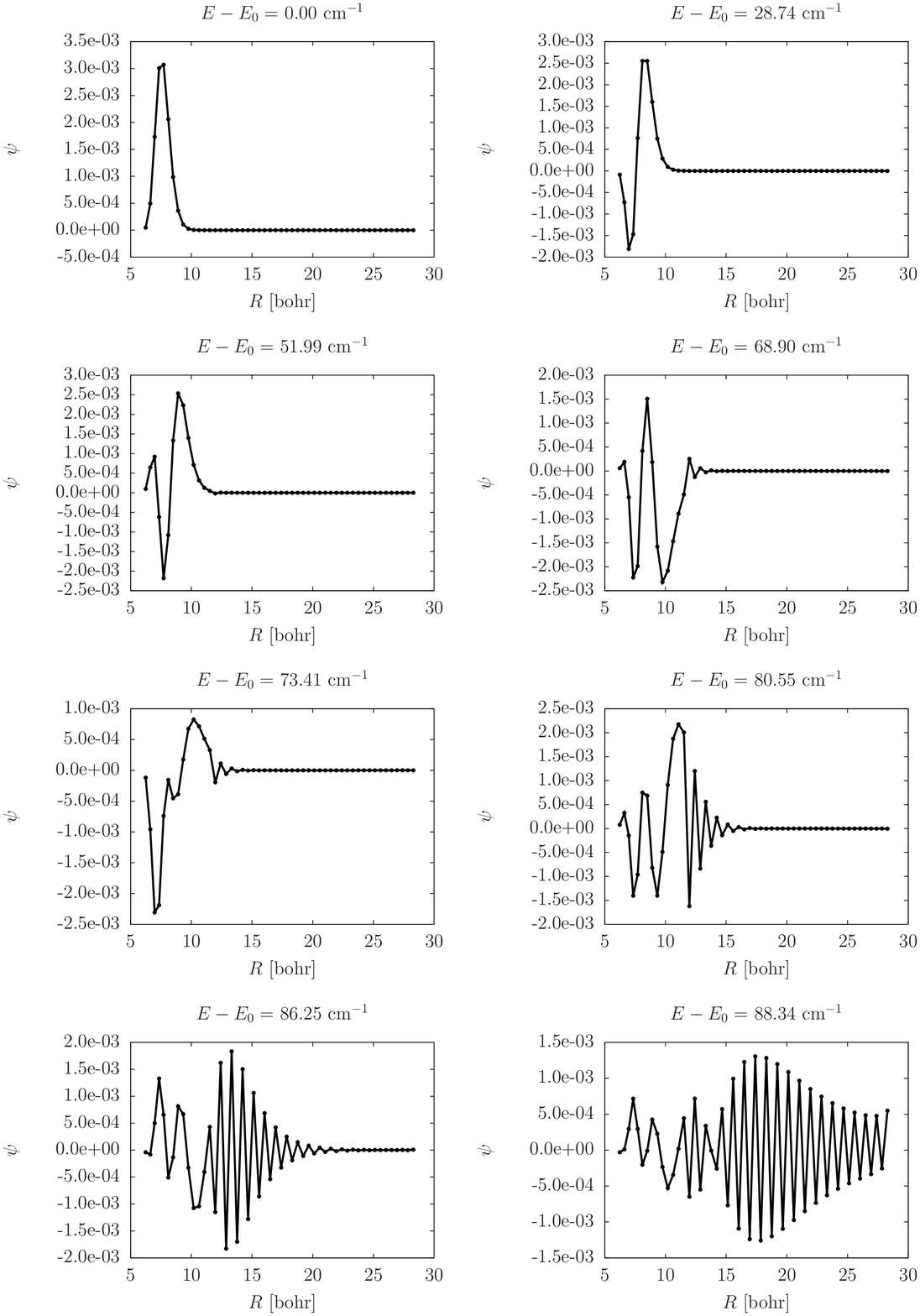}
\caption{Wave function cuts along the $R$ coordinate for the \aone-symmetry vibrational states.}
\label{figure06}
\end{figure}

\begin{figure}[H]
\begin{center}
\includegraphics[scale=1.3]{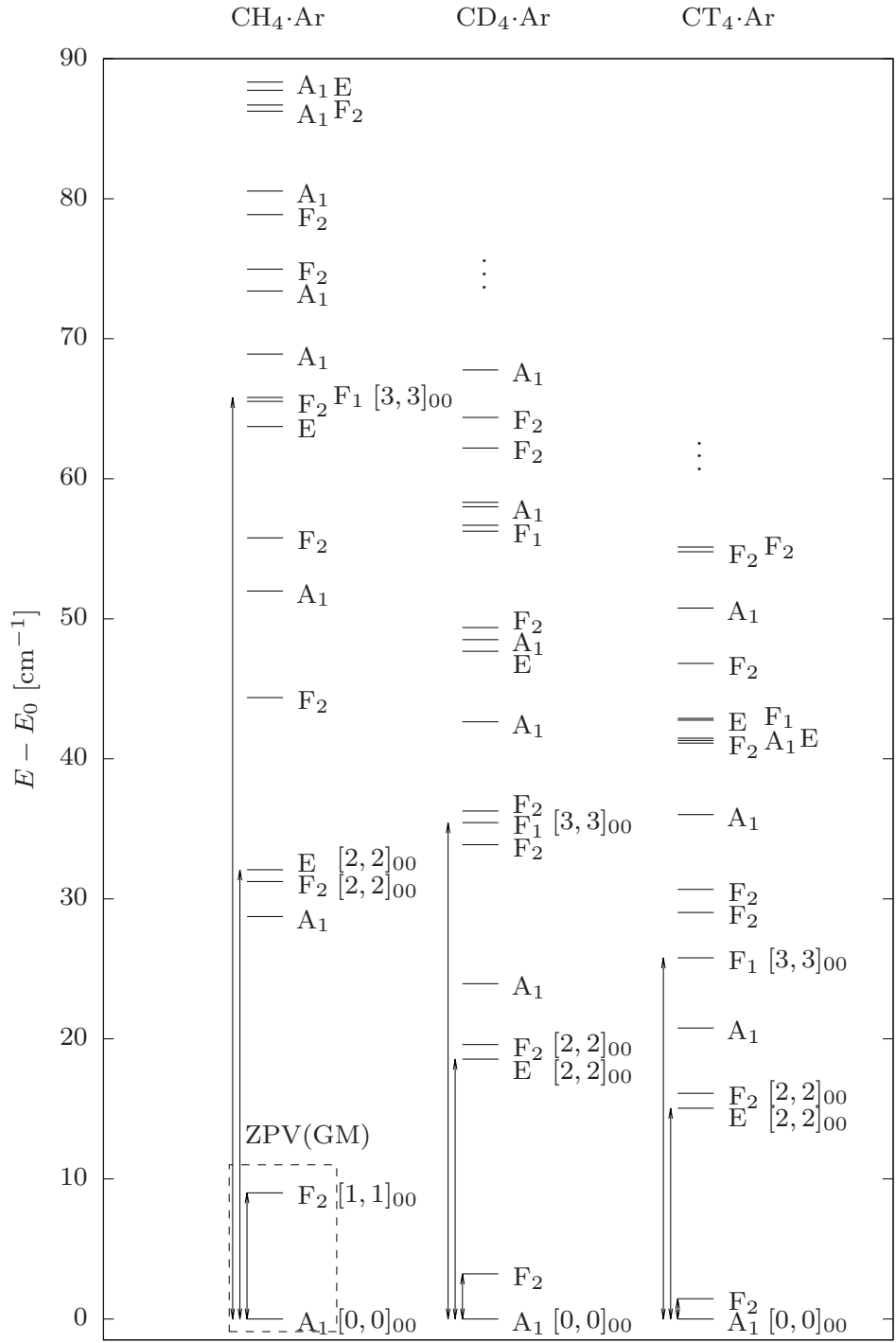}
\end{center}
\caption{%
  Comparison of the bound-state vibrational energies of CH$_4\cdot$Ar, CD$_4\cdot$Ar, and CT$_4\cdot$Ar. 
}
\label{figure07}
\end{figure}

Furthermore, the higher the energy, the number of 
nodes along $R$ is less obvious (Fig.~\ref{figure06}). While, we can trace 
(the A$_1$ species of)
the first, second, and perhaps also the third stretching excitations 
by node counting, beyond J0.26 (68.90~\cm, A$_1$),
the wave functions along $R$ become more and more oscillatory indicating
that we approach the dissociation limit.

It is interesting to note that there is not any bound vibrational state of A$_2$ symmetry, which would be an ``indicator'' of a vibrational state of A$_2$(GM) or an A$_2$(SM) PG symmetry, Eqs.~(\ref{c3vsplit}) and (\ref{c2vsplit}). 
Table~IV shows that the lowest-energy CR-functions, 
which generate an A$_2$-symmetry state belong to $[6,6]_{00}$.
These CR functions have $\sim$210~\cm\ energy, much larger than the 
dissociation energy, which explains the absence of bound states of A$_2$ symmetry.


While it is difficult to use the traditional picture of molecular vibrations,
the CR model appears to be useful and we could identify dominant CR functions for all vibrational states of methane-argon (see Figure~\ref{figure04} and Table~\ref{resoultsJ0}). 

Upon isotopic substitution, we could not really observe the closure of the splittings (since there are not any clear splittings apart from ZPV(GM)), 
but we observed another regularity with respect to the increase of the hydrogenic mass.
Based on this regularity, we will explain the change in the energy ordering of the symmetry species along the CH$_4\cdot$Ar--CD$_4\cdot$Ar--CT$_4\cdot$Ar sequence (Fig.~\ref{figure07}).

The rotational constant, and thus the rotational energy of methane is inversely 
proportional to the mass of the H (D/T) atom, and thus
the energy of the $[j^\text{M},\Lambda]_{00}$ CR functions is also
proportional to $m_\text{H}^{-1}$ to a good approximation 
(we neglect the energetic contribution from the diatom rotation, which is small).
If we check the energy separation of the lowest-energy vibrational states dominated 
by the $[0,0]_{00}$ (A$_1$), $[1,1]_{00}$ (F$_2$), $[2,2]_{00}$ (E$\oplus$F$_2$), 
and $[3,3]_{00}$ (A$_1\oplus$F$_1\oplus$F$_2$) CR functions, we observe
a ca. 1/2 and 1/3 reduction in the energy separation (indicated with double-headed arrows
in Fig.~\ref{figure07})
upon the
H--D and H--T replacement, respectively.
Of course, the changes in the vibrational energy intervals only approximately  
follow the CR model, because the PES also has
a role in the vibrational dynamics. 

In comparison with the $m_\text{H}^{-1}$ relation of the CR splittings, 
the energy of the atom-molecule stretching excitations (excitation along $R$)
changes approximately with $1/\sqrt{\mu}$, 
the square root of the reduced mass of the effective diatom (methane and argon),
which brings in a less than a $1/\sqrt{2}$ and $1/\sqrt{3}$ factor upon the H-D and H-T
replacements, respectively.

Hence, the energy separation of vibrational states corresponding to angular (CR) excitations (with similar radial parts)
is decreased by a larger extent than the energy separation of radial ($R$-stretching)
excitations upon isotopic substitution. Since radial excitation does not change the symmetry, while angular (CR) excitation does, we observe a rearrangement in the energy ordering of the various symmetry species in the deuterated and tritiated isotopologues.


\begin{table}
\caption{%
  Bound-state vibrational energies  
  given with respect to the $52.86$~\cm\ zero-point vibrational energy
  of \chfar\ obtained with GENIUSH and the PES of Ref. \cite{Kalugina}. 
  Each vibrational 
  state is characterized with its $T_{\textnormal{d}}$(M) irrep label 
  ($\mathit{\Gamma}$) 
  and the dominant (and minor) CR function.}
\begin{center}
\begin{tabular}{l c l c c}
\hline \hline
\textit{n}&$E$ [cm$^{-1}$]	&	$\mathit{\Gamma}$	&	\multicolumn{2}{c}{Coupled-rotor states} ($J=0$) \\ \cline{4-5}
 &       &           &     dominant         &    minor          \\
\hline 
J0.1	&	52.86	&	A$_1$	&	[0,0]$_{00}$	&				\\
J0.2--4	&	9.01	&	F$_2$	&	[1,1]$_{00}$	&				\\
J0.5	&	28.74	&	A$_1$	&	[0,0]$_{00}$	&				\\
J0.6--8	&	31.24	&	F$_2$	&	[2,2]$_{00}$	&	[1,1]$_{00}$			\\
J0.9--10	&	32.08	&	E	&	[2,2]$_{00}$	&				\\
J0.11--13	&	44.37	&	F$_2$	&	[1,1]$_{00}$	&	[2,2]$_{00}$			\\
J0.14	&	51.98	&	A$_1$	&	[0,0]$_{00}$	&				\\
J0.15--17	&	55.78	&	F$_2$	&	[2,2]$_{00}$	&	[1,1]$_{00}$	\&	[3,3]$_{00}$	\\
J0.18--19	&	63.73	&	E	&	[2,2]$_{00}$	&				\\
J0.20--22	&	65.53	&	F$_2$	&	[1,1]$_{00}$	&	[3,3]$_{00}$			\\
J0.23--25	&	65.81	&	F$_1$	&	[3,3]$_{00}$	&				\\
J0.26	&	68.90	&	A$_1$	&	[0,0]$_{00}$	&	[3,3]$_{00}$			\\
J0.27	&	73.41	&	A$_1$	&	[3,3]$_{00}$	&	[0,0]$_{00}$			\\
J0.28--30	&	74.97	&	F$_2$	&	[2,2]$_{00}$	&	[1,1]$_{00}$	\&	[3,3]$_{00}$	\\
J0.31--33	&	78.86	&	F$_2$	&	[1,1]$_{00}$	&	[2,2]$_{00}$			\\
J0.34	&	80.55	&	A$_1$	&	[0,0]$_{00}$	&				\\
J0.35	&	86.25	&	A$_1$	&	[0,0]$_{00}$	&				\\
J0.36--38	&	86.70	&	F$_2$	&	[1,1]$_{00}$	&	[2,2]$_{00}$	\&	[1,1]$_{00}$	\\
J0.39--40	&	87.74	&	E	&	[2,2]$_{00}$	&				\\
J0.41	&	88.33	&	A$_1$	&	[0,0]$_{00}$	&				\\

\hline \hline
\end{tabular}
\end{center}
\label{resoultsJ0}
\end{table}

%
%

\subsection[B]{Rovibrational states}

\begin{figure}
  \includegraphics[width=\textwidth]{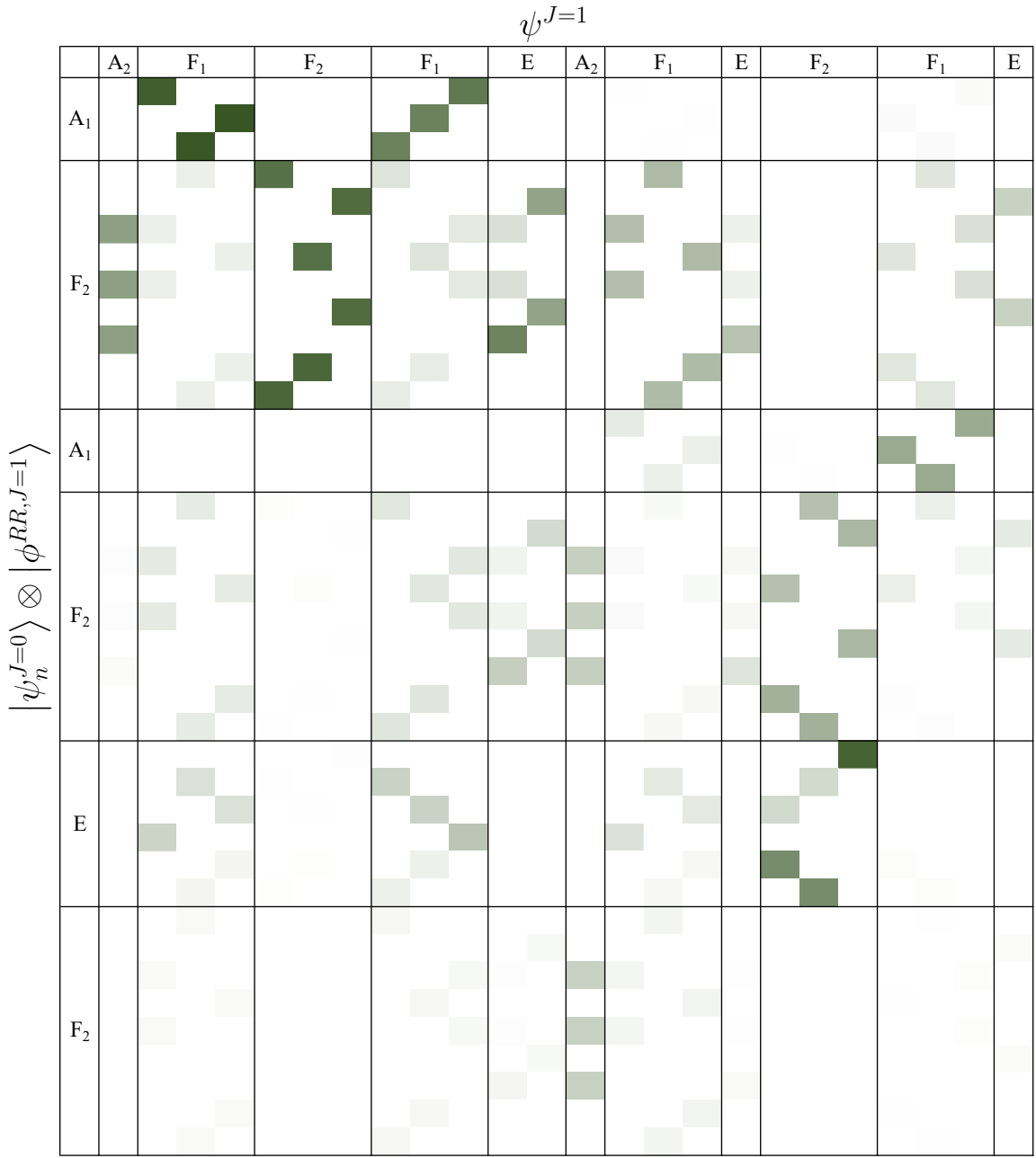}
  \caption{%
    RRD overlap matrix, Eq.~(\ref{RRDcoef}), for the $J=1$ rovibrational states of \chfar. The $\psi^{J=1}_n$ rovibrational functions listed in the columns are assigned with the product of the vibrational and rigid-rotor functions defined in Eq.~(\ref{RRDbasis}). 
  }
\label{RRDtablej1} 
\end{figure}

The computed $J=1$ rovibrational results are summarized in Table \ref{resoultsJ1}. The corresponding CRD matrix is shown in Fig.~\ref{figure04}, while the RRD coefficients (obtained with the RR functions corresponding to the symmetric-top GM structure) are visualized in Fig.~\ref{RRDtablej1}. A clear assignment, of the vibrational parent (with dominant overlaps) is possible only for the lower-energy range. Nevertheless, we find it remarkable that it is possible to use the RR picture at all for this very floppy complex, in the light of the difficulties of identifying complete vibrational splitting manifolds (see previous section).

\begin{figure}
\begin{center}
    \includegraphics[width=.9\columnwidth]{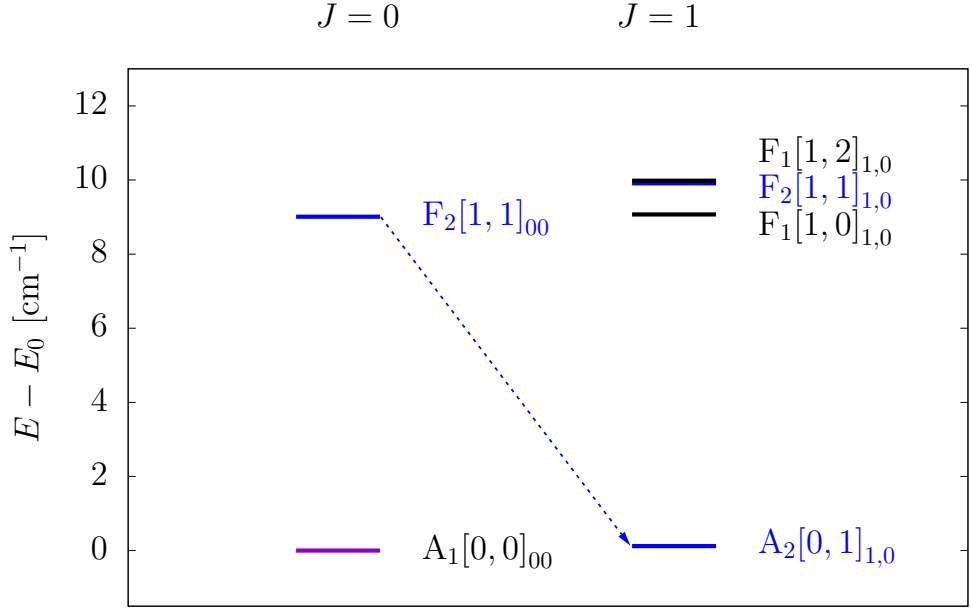}
\end{center}
\caption{%
  The lowest-energy $J=1$ rovibrational level of \chfar\ (0.12~\cm, \atwo)
  has lower energy than its vibrational parent state, \ftwo\ (9.01~\cm, \ftwo).
  The unusual energy ordering can be understood by inspecting the methane's rotational
  excitation in the dominant CR functions, $[0,1]_{1,0}$ and $[1,1]_{0,0}$, respectively.
}
\label{figure09} 
\end{figure}

Surprisingly, the very first $J=1$ rotationally excited state (0.12~\cm,  \atwo) has a lower energy than its vibrational parent state (9.01~\cm, \ftwo) highlighted in Fig.~\ref{figure09}.
The rotational excitation of the ZPV(GM) (0.00~\cm, \aone) appears only as the second rotational state
(9.07~\cm, \fone). 
This energy ordering of the rotational excitation of vibrational states is surprising if we have the traditional picture of rotating-vibrating molecules in mind. 
However, the traditional picture fails if the energy scale of rotations and intermolecular vibrations overlap which is the case for the present system.

Similarly to the vibrational states, the CR picture comes useful.
The dominant CR functions in the lowest energy rotational state and 
in its parent vibration are the $[0,1]_{10}$ and $[1,1]_{00}$ functions, respectively.
From this CR assignment, it can be seen that the unusual energy ordering is due to the rotational excitation of the methane unit which increases the energy but decreases the total rotational angular momentum, in the present case, 
due to its coupling with the rotating diatom.
The opposite situation is observed for the second-lowest rotational state and its parent vibration, in which 
the $[1,0]_{10}$ and $[0,0]_{00}$ are the dominant CR functions, respectively.
In this case, the rotational de-excitation of the methane molecule decreases the energy and (in this case) also the total rotational angular momentum is reduced.
The symmetry pattern of the states follows the formal regularities derived for the RR functions, Eq.~(\ref{rrdsym}) (see also Fig.~\ref{RRDtablej1}).

\begin{table}[h]
\caption{
  Bound-state rovibrational energies ($J=1$), measured from the zero-point vibration energy, of \chfar\ obtained with GENIUSH and the PES of Ref. \cite{Kalugina}. The assigned vibrational parent state, $T_{\textnormal{d}}$(M) irrep label ($\mathit{\Gamma}$), and the dominant (minor) coupled-rotor functions are also shown 
  for each rovibrational state.
}
\begin{center}
\resizebox{.77\textwidth}{!}{\begin{tabular}{l c c l c c}
\hline \hline
\textit{n} & $E$ [cm$^{-1}$]	 & \multicolumn{1}{c}{Vib. parent} 	 &	$\mathit{\Gamma}$	&	\multicolumn{2}{c}{Coupled-rotor states} ($J=1$) \\ \cline{5-6}
        &           &                   &           &     dominant      &    minor      \\
\hline 
J1.0	&	0.12	&	F$_2$ [9.01]	&	A$_2$	&	[0,1]$_{10}$	&				\\
J1.1--3	&	9.07	&	A$_1$ [0.00]	&	F$_1$	&	[1,0]$_{10}$	&	[1,2]$_{10}$			\\
J1.4--6	&	9.91	&	F$_2$ [9.01] \& E [32.08] 	&	F$_2$	&	[1,1]$_{10}$	&				\\
J1.7--9	&	9.98	&	A$_1$ [0.00]	&	F$_1$	&	[1,2]$_{10}$	&	[1,0]$_{10}$			\\
J1.10--11	&	22.57	&	F$_2$ [9.01] \& F$_2$ [31.24]	&	E	&	[2,2]$_{10}$	&	[2,1]$_{10}$	\&	[2,3]$_{10}$	\\
J1.12	&	28.86	&	F$_2$ [31.24] \& F$_2$  [44.37]	&	A$_2$	&	[0,1]$_{10}$	&				\\
J1.13--15	&	31.28	&	F$_2$ [9.01]	&	F$_1$	&	[2,1]$_{10}$	&	[1,2]$_{10}$			\\
J1.16--17	&	31.93	&	F$_2$ [9.01]	&	E	&	[2,1]$_{10}$	&	[2,3]$_{10}$			\\
J1.18--20	&	31.94	&	E [32.08] \& F$_2$ [31.24]	&	F$_2$	&	[2,2]$_{10}$	&	[1,1]$_{10}$			\\
J1.21--23	&	32.01	&	A$_1$ [28.74]	&	F$_1$	&	[2,3]$_{10}$	&	[1,0]$_{10}$	\&	[2,1]$_{10}$	\\
J1.24--26	&	40.45	&	F$_2$ [44.37]	&	F$_2$	&	[1,1]$_{10}$	&	[2,2]$_{10}$			\\
J1.27--29	&	40.46	&	A$_1$ [28.74]	&	F$_1$	&	[1,0]$_{10}$	&	[1,2]$_{10}$			\\
J1.30--31	&	42.82	&	F$_2$ [31.24]	&	E	&	[2,2]$_{10}$	&	[2,1]$_{10}$	\&	[2,3]$_{10}$	\\
J1.32--34	&	44.48	&	E [63.73]	&	F$_1$	&	[1,2]$_{10}$	&	[2,3]$_{10}$	\&	[1,0]$_{10}$	\\
J1.35	&	52.08	&	F$_2$ [55.78] \& F$_2$ [65.53]	&	A$_2$	&	[0,1]$_{10}$	&				\\
J1.36--37	&	55.69	&	F$_2$ [44.37]	&	E	&	[2,2]$_{10}$	&	[2,1]$_{10}$			\\
J1.38--40	&	55.84	&	F$_2$ [31.24]	&	F$_1$	&	[2,3]$_{10}$	&	[1,2]$_{10}$	\&	[2,1]$_{10}$	\\
J1.41--43	&	58.32	&	F$_2$ [55.78]	&	F$_2$	&	[1,1]$_{10}$	&	[2,2]$_{10}$			\\
J1.44--46	&	58.34	&	F$_1$ [51.98]	&	F$_1$	&	[1,0]$_{10}$	&	[3,2]$_{10}$	\&	[1,2]$_{10}$	\\
J1.47--48	&	63.58	&	F$_2$ [44.37]	&	E	&	[2,3]$_{10}$	&	[2,1]$_{10}$			\\
J1.49--51	&	65.55	&		&	F$_1$	&	[1,2]$_{10}$	&	[3,2]$_{10}$			\\
J1.52	&	65.78	&	F$_2$ [31.24] \& F$_2$  [44.37]	&	A$_2$	&	[3,2]$_{10}$	&	[3,4]$_{10}$			\\
J1.53	&	65.83	&	F$_1$ [65.81]	&	A$_1$	&	[3,3]$_{10}$	&				\\
J1.54--56	&	65.91	&	F$_2$ [65.53] \& F$_1$ [65.81]	&	F$_2$	&	[1,1]$_{10}$	&	[3,2]$_{10}$			\\
J1.57--59	&	66.03	&	A$_1$ [51.98]	&	F$_1$	&	[1,0]$_{10}$	&	[3,4]$_{10}$			\\
J1.60--62	&	66.08	&	F$_2$ [65.53]	&	F$_2$	&	[3,4]$_{10}$	&	[1,1]$_{10}$			\\
J1.63	&	68.99	&	F$_2$ [78.86]	&	A$_2$	&	[0,1]$_{10}$	&				\\
J1.64--65	&	71.49	&	F$_2$ [55.78] \& F$_2$ [74.97]	&	E	&	[2,2]$_{10}$	&	[2,3]$_{10}$	\&	[2,1]$_{10}$	\\
J1.66	&	73.58	&	F$_2$ [44.37]	&	A$_2$	&	[3,4]$_{10}$	&	[3,2]$_{10}$	\&	[0,1]$_{10}$	\\
J1.67--69	&	74.33	&		&	F$_1$	&	[3,3]$_{10}$	&	[3,2]$_{10}$	\&	[2,1]$_{10}$	\\
J1.70--72	&	74.42	&		&	F$_2$	&	[3,2]$_{10}$	&	[3,3]$_{10}$	\&	[3,4]$_{10}$	\\
J1.73--75	&	75.12	&		&	F$_1$	&	[2,3]$_{10}$	&	[3,4]$_{10}$			\\
J1.76--78	&	78.95	&	A$_1$ [68.90]	&	F$_1$	&	[1,2]$_{10}$	&	[1,0]$_{10}$	\&	[2,3]$_{10}$	\\
J1.79--81	&	79.54	&	F$_2$ [78.86]	&	F$_2$	&	[1,1]$_{10}$	&				\\
J1.82--84	&	79.55	&	A$_1$ [68.90]	&	F$_1$	&	[1,0]$_{10}$	&	[1,2]$_{10}$			\\
J1.85	&	80.60	&		&	A$_2$	&	[0,1]$_{10}$	&				\\
J1.86--87	&	81.61	&	Several F$_2$ states	&	E	&	[2,2]$_{10}$	&	[2,1]$_{10}$	\&	[2,3]$_{10}$	\\
J1.88	&	86.26	&		&	A$_2$	&	[0,1]$_{10}$	&				\\
J1.89--91	&	86.69	&		&	F$_1$	&	[1,2]$_{10}$	&	[2,1]$_{10}$	\&	[1,0]$_{10}$	\\
J1.92--94	&	87.31	&	F$_2$ [86.70]  \& E [87.74]	&	F$_2$	&	[2,2]$_{10}$	&	[1,1]$_{10}$			\\
J1.95--97	&	87.34	&	A$_1$ [80.55]	&	F$_1$	&	[2,3]$_{10}$	&	[2,1]$_{10}$	\&	[1,0]$_{10}$	\\
J1.98--99	&	87.57	&	F$_2$ [65.53]	&	E	&	[2,3]$_{10}$	&	[2,1]$_{10}$			\\
J1.100	&	88.31	&		&	A$_2$	&	[0,1]$_{10}$	&				\\
\hline \hline
\end{tabular}} 
\end{center}
\label{resoultsJ1}
\end{table}

%
%

\subsection[C]{Resonance states}
\begin{figure}
\begin{center}
\includegraphics[width=0.7\textwidth]{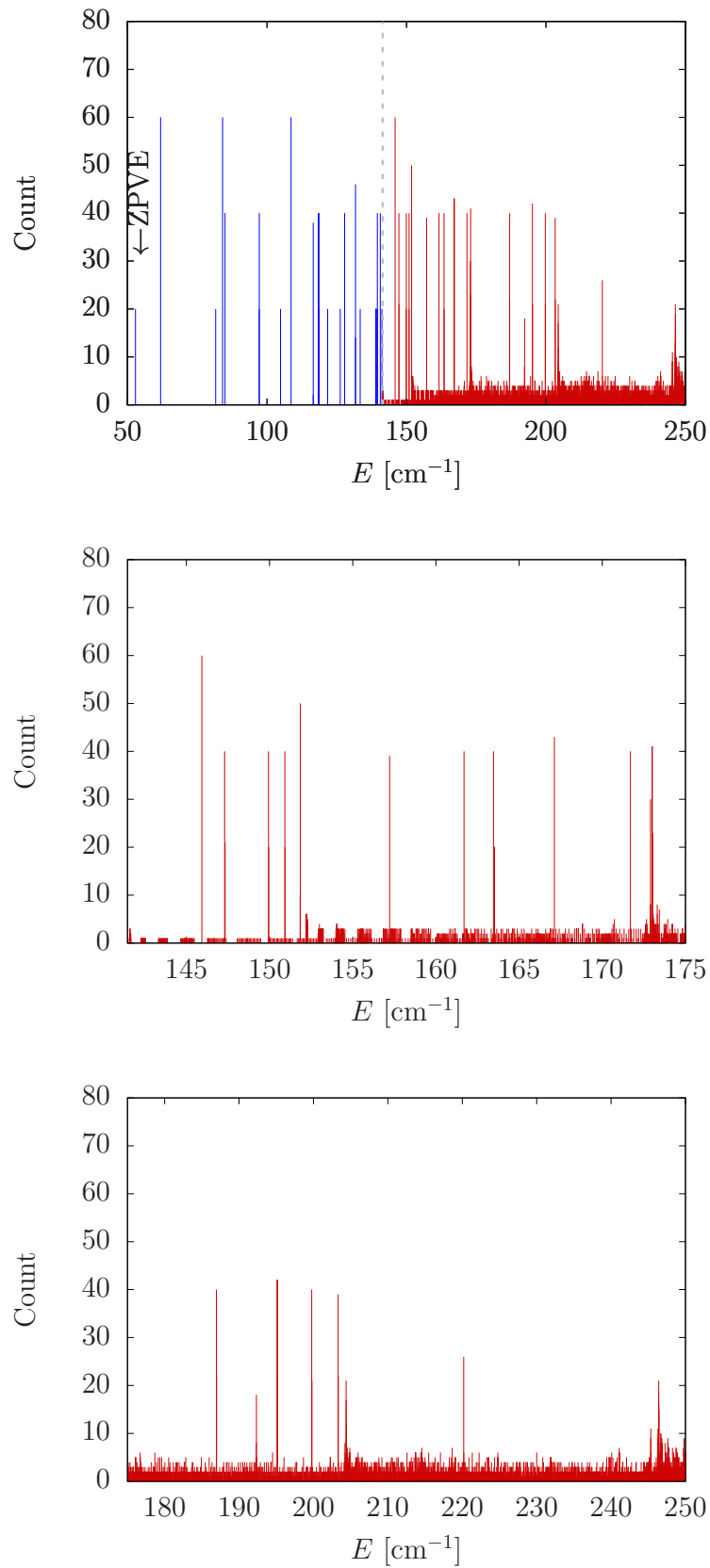}
\end{center}
\caption{
  Bound (blue) and resonance (red) vibrational states of \chfar\ 
  indicated by the peaks in the stabilization histogram. measured from
  the global minimum of the PES. (For some triply degenerate states an artificial split can be observed in the histogram due to imperfect convergence. An unambiguous symmetry assignment is established through a CRD analysis (see Table~\ref{tab:res}).)}
\label{figure10}
\end{figure}

Twenty vibrational bound states have been identified up to the dissociation 
asymptote (Table~\ref{resoultsJ0}), 
and we continued the computations beyond the dissociation limit.
Repeated computations with slightly different grid intervals of the 
molecule-atom distance allowed us to 
identify long-lived resonance states using the stabilization method \cite{stabilization} and a histogram analysis technique \cite{ArNOp}. 
Twenty GENIUSH computations were carried out, and 
1000 eigenvalues were computed in each run with different box sizes. 
In every computation the maximal value for the $R$ coordinate was slightly modified, from 14.05 \r{A} to 15 \r{A}
(the PES is defined up to $R=10$~\r{A}), incremented by 0.05 \r{A} in each computation. The computed eigenvalues are collected and sorted from the lowest to the largest, then a histogram is made from the data. The eigenvalues corresponding to the resonance states will appear in every computation, meanwhile the continuous eigenvalues are shifted by changing the size of the box. The obtained histogram (bin size 0.01~cm$^{-1}$) is shown in Figure~\ref{figure10}
and the energy positions and assignment for the peaks are collected in Table~\ref{tab:res}. 

It is interesting to note that we find only E and \ftwo-symmetry states among the identified long-lived resonances. 
The dominant CR functions in these states are $[1,1]_{00}$, $[2,2]_{00}$, and $[3,3]_{00}$.
The lack of A$_1$ states may be explained by the fact that highly excited vibrations along the $R$ (dissociation) coordinate also have A$_1$ symmetry, and due to the coupling of states with the same symmetry, 
we do not observed any long-lived resonance of this species.

The absence of states dominated  by the $[0,0]_{00}$ CR function 
tells us that a fair portion of the energy in these long-lived resonances 
appears in the angular degrees of freedom
(the excitation of which is measured by the CRD).
The presumably slow energy transfer from the angular degrees of freedom to the 
radial (dissociative) degree of freedom gives rise to a long lifetime of the identified states.

It is also insightful to compare the energy of the states with 
the dissociation energy $D_0(\text{GM})\approx 90$~\cm\ (measured from the ZPV as in Fig.~10) 
and the CR energies of the $[n,n]_{00}$ functions with $n=1,2,$ and $3$: ca.~0~\cm, 10~\cm, 30~\cm, and 60~\cm, respectively. Assuming simple additivity of these energies, we may understand
that long-lived states dominated by $[1,1]_{00}$ are possible up to ca.~100~\cm\ (however, we do not have
any simple explanation for the J0.73--75, F$_2$ state).
Similarly, we may find long-lived states dominated by $[2,2]_{00}$ or $[3,3]_{00}$ 
up to ca.~130--140~\cm, which is indeed the case.

All peaks above the $D_e$ dissociation energy (141.47 cm$^{-1}$) correspond to a resonance state (Fig. \ref{figure10}.). The symmetry assignment by CRD and the major and minor coupled rotor contributions are listed in Table \ref{resonanceJ0}. 

\begin{table}
\caption{
  Energy of the long-lived ($J=0$) vibrational resonances of \chfar\ computed with GENIUSH, the PES of Ref. \cite{Kalugina}, and the stabilization technique. The vibrational states are given with respect to the zero-point vibration energy. Each state is characterized with its $T_{\textnormal{d}}$(M) irrep label and the dominant (and minor) coupled-rotor functions.
\label{tab:res}}
\begin{center}
\begin{tabular}{l c l c c}
\hline \hline
\textit{n}&$E$ [cm$^{-1}$]	&	$\mathit{\Gamma}$	&	\multicolumn{2}{c}{Coupled-rotor states} ($J=0$) \\ \cline{4-5}
 &       &           &     dominant         &    minor          \\
\hline 
J0.42--44	&	93.06	&	F$_2$	&	[1,1]$_{00}$	&		\\
J0.45--47	&	94.44	&	F$_2$	&	[3,3]$_{00}$	&	[2,2]$_{00}$	\\
J0.48--50	&	97.08	&	F$_2$	&	[1,1]$_{00}$	&	[2,2]$_{00}$	\\
J0.51--52	&	98.06	&	F$_2$	&	[1,1]$_{00}$	&	[2,2]$_{00}$	\\
J0.53--55	&	98.98	&	F$_2$	&	[1,1]$_{00}$	&		\\
J0.56--57	&	104.36	&	E	&	[2,2]$_{00}$	&		\\
J0.58--59	&	108.82	&	E	&	[3,3]$_{00}$	&		\\
J0.60--62	&	110.60	&	F$_2$	&	[3,3]$_{00}$	&	[2,2]$_{00}$	\\
J0.63--64	&	114.25	&	E	&	[2,2]$_{00}$	&		\\
J0.65--66	&	118.82	&	E	&	[2,2]$_{00}$	&		\\
J0.67--68	&	120.03	&	E	&	[2,2]$_{00}$	&		\\
J0.69--70	&	134.08	&	E	&	[2,2]$_{00}$	&	[3,3]$_{00}$	\\
J0.71--72	&	139.46	&	E	&	[3,3]$_{00}$	&		\\
J0.73--75	&	142.34	&	F$_2$	&	[1,1]$_{00}$	&	[2,2]$_{00}$	\\
J0.76--78	&	150.45	&	F$_2$	&	[3,3]$_{00}$	&	[2,2]$_{00}$	\\
\hline \hline
\end{tabular}
\end{center}
\label{resonanceJ0}
\end{table}

%
%
\section{Summary}
Rovibrational computations have been carried out for the methane-argon dimer 
using a new, \emph{ab initio} atom-molecule potential energy surface \cite{Kalugina}
and the general rovibrational program, GENIUSH \cite{geniush,geniush2}. 
This PES, with a well depth of $-D_\text{eq}=-141.47$~\cm, 
supports twenty bound states (counting degenerate states only once).
The energy of the lowest accessible vibrational state to the system is 
$-D_0(GM)=-88.60$~\cm, measured from the dissociation asymptote. 
Although there are two different minimum structures,
with a  26.41~\cm\ energy separation, 
most of the computed (ro)vibraitonal states cannot be unambiguously assigned 
either to the global or to the secondary minimum.
The computed rovibrational states
were analyzed using two limiting models and detailed symmetry considerations.

The traditional picture of molecular rotations and vibrations almost completely breaks down
for this weakly bound, floppy system. We could identify the complete tunneling splitting only for
the zero-point vibrational state of the global minimum. Vibrational parent states could
be determined only for the lowest few rotational states with $J=1$ rotational quantum number.
Interestingly, the lowest-energy rotational state does not correspond to the rotational excitation of 
lowest-energy vibrational state
but it corresponds to the second lowest vibrational state. 
This rotational state and its vibrational parent 
represent a ``formally'' negative-energy rotational excitation, \emph{i.e.,}
the rotationally excited state has a lower energy than its parent vibration.

These unusual features, which appear strange in terms of the traditional picture of rotating-vibrating molecules,
can be well understood using a model which considers the rotating-vibrating complex
in terms of coupled quantum rotors of rigidly rotating methane and the effective diatom (atom-molecule separation). 
All vibrational bound and 
low-energy, long-lived resonance states as well as most of the $J=1$ rovibrational states can be assigned
within this coupled-rotor picture.

%
%

\section{Acknowledgement}
\noindent %
We thank Yulia Kalugina for sending to us the potential energy subroutine
of methane-argon, and Gustavo Avila for initial discussions on this project.
We acknowledge financial support from a PROMYS Grant (no. IZ11Z0\_166525)  
of the Swiss National Science Foundation.
F.D. also thanks the New National Excellence Program of 
the Ministry of Human Capacities of Hungary
(ÚNKP-18-3-II-ELTE-133). 
A PRACE Preparatory Access Grant allowed us to test the applicability
and scalability of the GENIUSH program on European Supercomputers (Barcelona, Paris, and Stuttgart), which is gratefully acknowledged.

\clearpage

\begin{thebibliography}{24}
\expandafter\ifx\csname natexlab\endcsname\relax\def\natexlab#1{#1}\fi
\expandafter\ifx\csname bibnamefont\endcsname\relax
  \def\bibnamefont#1{#1}\fi
\expandafter\ifx\csname bibfnamefont\endcsname\relax
  \def\bibfnamefont#1{#1}\fi
\expandafter\ifx\csname citenamefont\endcsname\relax
  \def\citenamefont#1{#1}\fi
\expandafter\ifx\csname url\endcsname\relax
  \def\url#1{\texttt{#1}}\fi
\expandafter\ifx\csname urlprefix\endcsname\relax\def\urlprefix{URL }\fi
\providecommand{\bibinfo}[2]{#2}
\providecommand{\eprint}[2][]{\url{#2}}

\bibitem[{\citenamefont{Kalugina et~al.}(2016)\citenamefont{Kalugina,
  Lokshtanov, Cherepanov, and Vigasin}}]{Kalugina}
\bibinfo{author}{\bibfnamefont{Y.~N.} \bibnamefont{Kalugina}},
  \bibinfo{author}{\bibfnamefont{S.~E.} \bibnamefont{Lokshtanov}},
  \bibinfo{author}{\bibfnamefont{V.~N.} \bibnamefont{Cherepanov}},
  \bibnamefont{and} \bibinfo{author}{\bibfnamefont{A.~A.}
  \bibnamefont{Vigasin}}, \bibinfo{journal}{J. Chem. Phys.}
  \textbf{\bibinfo{volume}{144}}, \bibinfo{pages}{054304}
  (\bibinfo{year}{2016}).

\bibitem[{\citenamefont{Liu and Jäger}(2004)}]{Jager}
\bibinfo{author}{\bibfnamefont{Y.}~\bibnamefont{Liu}} \bibnamefont{and}
  \bibinfo{author}{\bibfnamefont{W.}~\bibnamefont{Jäger}},
  \bibinfo{journal}{J. Chem. Phys.} \textbf{\bibinfo{volume}{120}},
  \bibinfo{pages}{9047} (\bibinfo{year}{2004}).

\bibitem[{\citenamefont{Samuelson et~al.}(1997)\citenamefont{Samuelson, Nath,
  and Borysow}}]{SAMUELSON}
\bibinfo{author}{\bibfnamefont{R.~E.} \bibnamefont{Samuelson}},
  \bibinfo{author}{\bibfnamefont{N.~R.} \bibnamefont{Nath}}, \bibnamefont{and}
  \bibinfo{author}{\bibfnamefont{A.}~\bibnamefont{Borysow}},
  \bibinfo{journal}{Planetary and Space Science} \textbf{\bibinfo{volume}{45}},
  \bibinfo{pages}{959 } (\bibinfo{year}{1997}), ISSN \bibinfo{issn}{0032-0633}.

\bibitem[{\citenamefont{Sarka et~al.}(2016)\citenamefont{Sarka, Csaszar,
  Althorpe, Wales, and Matyus}}]{ch4h2o2016}
\bibinfo{author}{\bibfnamefont{J.}~\bibnamefont{Sarka}},
  \bibinfo{author}{\bibfnamefont{A.~G.} \bibnamefont{Csaszar}},
  \bibinfo{author}{\bibfnamefont{S.~C.} \bibnamefont{Althorpe}},
  \bibinfo{author}{\bibfnamefont{D.~J.} \bibnamefont{Wales}}, \bibnamefont{and}
  \bibinfo{author}{\bibfnamefont{E.}~\bibnamefont{Matyus}},
  \bibinfo{journal}{Phys. Chem. Chem. Phys.} \textbf{\bibinfo{volume}{18}},
  \bibinfo{pages}{22816} (\bibinfo{year}{2016}).

\bibitem[{\citenamefont{Sarka et~al.}(2017)\citenamefont{Sarka, Csaszar, and
  Matyus}}]{ch4h2o}
\bibinfo{author}{\bibfnamefont{J.}~\bibnamefont{Sarka}},
  \bibinfo{author}{\bibfnamefont{A.~G.} \bibnamefont{Csaszar}},
  \bibnamefont{and} \bibinfo{author}{\bibfnamefont{E.}~\bibnamefont{Matyus}},
  \bibinfo{journal}{Phys. Chem. Chem. Phys.} \textbf{\bibinfo{volume}{19}},
  \bibinfo{pages}{15335} (\bibinfo{year}{2017}).

\bibitem[{\citenamefont{Fábri et~al.}(2013)\citenamefont{Fábri, Császár,
  and Czakó}}]{ch4f-}
\bibinfo{author}{\bibfnamefont{C.}~\bibnamefont{Fábri}},
  \bibinfo{author}{\bibfnamefont{A.~G.} \bibnamefont{Császár}},
  \bibnamefont{and} \bibinfo{author}{\bibfnamefont{G.}~\bibnamefont{Czakó}},
  \bibinfo{journal}{J. Phys. Chem. A} \textbf{\bibinfo{volume}{117}},
  \bibinfo{pages}{6975} (\bibinfo{year}{2013}).

\bibitem[{\citenamefont{Buryak et~al.}(2013)\citenamefont{Buryak, Kalugina, and
  Vigasin}}]{ch4ch4}
\bibinfo{author}{\bibfnamefont{I.}~\bibnamefont{Buryak}},
  \bibinfo{author}{\bibfnamefont{Y.}~\bibnamefont{Kalugina}}, \bibnamefont{and}
  \bibinfo{author}{\bibfnamefont{A.}~\bibnamefont{Vigasin}},
  \bibinfo{journal}{J. Mol. Spectrosc.} \textbf{\bibinfo{volume}{291}},
  \bibinfo{pages}{102 } (\bibinfo{year}{2013}).

\bibitem[{\citenamefont{Dore and Filabozzi}(1990)}]{filabozzi}
\bibinfo{author}{\bibfnamefont{P.}~\bibnamefont{Dore}} \bibnamefont{and}
  \bibinfo{author}{\bibfnamefont{A.}~\bibnamefont{Filabozzi}},
  \bibinfo{journal}{Can. J. Phys.} \textbf{\bibinfo{volume}{68}},
  \bibinfo{pages}{1196} (\bibinfo{year}{1990}).

\bibitem[{\citenamefont{McKellar}(1994)}]{mckellar}
\bibinfo{author}{\bibfnamefont{A.~R.~W.} \bibnamefont{McKellar}},
  \bibinfo{journal}{Faraday Discuss.} \textbf{\bibinfo{volume}{97}},
  \bibinfo{pages}{69} (\bibinfo{year}{1994}).

\bibitem[{\citenamefont{Wangler et~al.}(2003)\citenamefont{Wangler, Roth, Pak,
  Winnewisser, Wormer, and van~der Avoird}}]{WANGLER}
\bibinfo{author}{\bibfnamefont{M.}~\bibnamefont{Wangler}},
  \bibinfo{author}{\bibfnamefont{D.}~\bibnamefont{Roth}},
  \bibinfo{author}{\bibfnamefont{I.}~\bibnamefont{Pak}},
  \bibinfo{author}{\bibfnamefont{G.}~\bibnamefont{Winnewisser}},
  \bibinfo{author}{\bibfnamefont{P.}~\bibnamefont{Wormer}}, \bibnamefont{and}
  \bibinfo{author}{\bibfnamefont{A.}~\bibnamefont{van~der Avoird}},
  \bibinfo{journal}{J. Mol. Spectrosc.} \textbf{\bibinfo{volume}{222}},
  \bibinfo{pages}{109 } (\bibinfo{year}{2003}).

\bibitem[{\citenamefont{Mátyus et~al.}(2009)\citenamefont{Mátyus, Czakó, and
  Császár}}]{geniush}
\bibinfo{author}{\bibfnamefont{E.}~\bibnamefont{Mátyus}},
  \bibinfo{author}{\bibfnamefont{G.}~\bibnamefont{Czakó}}, \bibnamefont{and}
  \bibinfo{author}{\bibfnamefont{A.~G.} \bibnamefont{Császár}},
  \bibinfo{journal}{J. Chem. Phys.} \textbf{\bibinfo{volume}{130}},
  \bibinfo{pages}{134112} (\bibinfo{year}{2009}).

\bibitem[{\citenamefont{Fábri et~al.}(2011)\citenamefont{Fábri, Mátyus, and
  Császár}}]{geniush2}
\bibinfo{author}{\bibfnamefont{C.}~\bibnamefont{Fábri}},
  \bibinfo{author}{\bibfnamefont{E.}~\bibnamefont{Mátyus}}, \bibnamefont{and}
  \bibinfo{author}{\bibfnamefont{A.~G.} \bibnamefont{Császár}},
  \bibinfo{journal}{J. Chem. Phys.} \textbf{\bibinfo{volume}{134}},
  \bibinfo{pages}{074105} (\bibinfo{year}{2011}).

\bibitem[{\citenamefont{Papp et~al.}(2017{\natexlab{a}})\citenamefont{Papp,
  Sarka, Szidarovszky, Császár, Mátyus, Hochlaf, and Stoecklin}}]{ArNOp}
\bibinfo{author}{\bibfnamefont{D.}~\bibnamefont{Papp}},
  \bibinfo{author}{\bibfnamefont{J.}~\bibnamefont{Sarka}},
  \bibinfo{author}{\bibfnamefont{T.}~\bibnamefont{Szidarovszky}},
  \bibinfo{author}{\bibfnamefont{A.~G.} \bibnamefont{Császár}},
  \bibinfo{author}{\bibfnamefont{E.}~\bibnamefont{Mátyus}},
  \bibinfo{author}{\bibfnamefont{M.}~\bibnamefont{Hochlaf}}, \bibnamefont{and}
  \bibinfo{author}{\bibfnamefont{T.}~\bibnamefont{Stoecklin}},
  \bibinfo{journal}{Phys. Chem. Chem. Phys.} \textbf{\bibinfo{volume}{19}},
  \bibinfo{pages}{8152} (\bibinfo{year}{2017}{\natexlab{a}}).

\bibitem[{\citenamefont{Papp et~al.}(2017{\natexlab{b}})\citenamefont{Papp,
  Szidarovszky, and Császár}}]{H2He}
\bibinfo{author}{\bibfnamefont{D.}~\bibnamefont{Papp}},
  \bibinfo{author}{\bibfnamefont{T.}~\bibnamefont{Szidarovszky}},
  \bibnamefont{and} \bibinfo{author}{\bibfnamefont{A.~G.}
  \bibnamefont{Császár}}, \bibinfo{journal}{J. Chem. Phys.}
  \textbf{\bibinfo{volume}{147}}, \bibinfo{pages}{094106}
  (\bibinfo{year}{2017}{\natexlab{b}}).

\bibitem[{\citenamefont{Fábri et~al.}(2014)\citenamefont{Fábri, Sarka, and
  Császár}}]{h5+}
\bibinfo{author}{\bibfnamefont{C.}~\bibnamefont{Fábri}},
  \bibinfo{author}{\bibfnamefont{J.}~\bibnamefont{Sarka}}, \bibnamefont{and}
  \bibinfo{author}{\bibfnamefont{A.~G.} \bibnamefont{Császár}},
  \bibinfo{journal}{J. Chem. Phys.} \textbf{\bibinfo{volume}{140}},
  \bibinfo{pages}{051101} (\bibinfo{year}{2014}).

\bibitem[{\citenamefont{Fábri et~al.}(2017)\citenamefont{Fábri, Quack, and
  Császár}}]{FaQuCs17}
\bibinfo{author}{\bibfnamefont{C.}~\bibnamefont{Fábri}},
  \bibinfo{author}{\bibfnamefont{M.}~\bibnamefont{Quack}}, \bibnamefont{and}
  \bibinfo{author}{\bibfnamefont{A.~G.} \bibnamefont{Császár}},
  \bibinfo{journal}{J. Chem. Phys.} \textbf{\bibinfo{volume}{147}},
  \bibinfo{pages}{134101} (\bibinfo{year}{2017}).

\bibitem[{\citenamefont{Wang and Jr.}(2016)}]{ch5+}
\bibinfo{author}{\bibfnamefont{X.-G.} \bibnamefont{Wang}} \bibnamefont{and}
  \bibinfo{author}{\bibfnamefont{T.~C.} \bibnamefont{Jr.}},
  \bibinfo{journal}{J. Chem. Phys.} \textbf{\bibinfo{volume}{144}},
  \bibinfo{pages}{204304} (\bibinfo{year}{2016}).

\bibitem[{\citenamefont{Schmiedt et~al.}(2015)\citenamefont{Schmiedt,
  Schlemmer, and Jensen}}]{ch5+2}
\bibinfo{author}{\bibfnamefont{H.}~\bibnamefont{Schmiedt}},
  \bibinfo{author}{\bibfnamefont{S.}~\bibnamefont{Schlemmer}},
  \bibnamefont{and} \bibinfo{author}{\bibfnamefont{P.}~\bibnamefont{Jensen}},
  \bibinfo{journal}{J. Chem. Phys.} \textbf{\bibinfo{volume}{143}},
  \bibinfo{pages}{154302} (\bibinfo{year}{2015}).

\bibitem[{\citenamefont{Wodraszka and Manthe}(2015)}]{ch5+3}
\bibinfo{author}{\bibfnamefont{R.}~\bibnamefont{Wodraszka}} \bibnamefont{and}
  \bibinfo{author}{\bibfnamefont{U.}~\bibnamefont{Manthe}},
  \textbf{\bibinfo{volume}{6}}, \bibinfo{pages}{4229} (\bibinfo{year}{2015}).

\bibitem[{\citenamefont{Coursey et~al.}(2015)\citenamefont{Coursey, Schwab,
  Tsai, and Dragoset}}]{NIST}
\bibinfo{author}{\bibfnamefont{J.~S.} \bibnamefont{Coursey}},
  \bibinfo{author}{\bibfnamefont{D.~J.} \bibnamefont{Schwab}},
  \bibinfo{author}{\bibfnamefont{J.~J.} \bibnamefont{Tsai}}, \bibnamefont{and}
  \bibinfo{author}{\bibfnamefont{R.~A.} \bibnamefont{Dragoset}},
  \bibinfo{howpublished}{{Atomic Weights and Isotopic Compositions (version
  4.1): {\tt{http://physics.nist.gov/Comp}} [last accessed on 12 May 2018].
  National Institute of Standards and Technology, Gaithersburg, MD.}}
  (\bibinfo{year}{2015}).

\bibitem[{\citenamefont{Bunker and Jensen}(2006)}]{bunker-jensen}
\bibinfo{author}{\bibfnamefont{P.}~\bibnamefont{Bunker}} \bibnamefont{and}
  \bibinfo{author}{\bibfnamefont{P.}~\bibnamefont{Jensen}},
  \emph{\bibinfo{title}{Molecular {S}ymmetry and {S}pectroscopy}}
  (\bibinfo{publisher}{NRC Research Press}, \bibinfo{year}{2006}).

\bibitem[{\citenamefont{Wilson et~al.}(1980)\citenamefont{Wilson, Decius, and
  Cross}}]{wilson}
\bibinfo{author}{\bibfnamefont{E.~B.} \bibnamefont{Wilson}},
  \bibinfo{author}{\bibfnamefont{J.~C.} \bibnamefont{Decius}},
  \bibnamefont{and} \bibinfo{author}{\bibfnamefont{P.~C.} \bibnamefont{Cross}},
  \emph{\bibinfo{title}{Molecular {V}ibrations: the {T}heory of {I}nfrared and
  {R}aman {V}ibrational {S}pectra}} (\bibinfo{publisher}{Courier Corporation},
  \bibinfo{year}{1980}).

\bibitem[{\citenamefont{Mátyus et~al.}(2010)\citenamefont{Mátyus, Fábri,
  Szidarovszky, Czakó, Allen, and Császár}}]{RRD}
\bibinfo{author}{\bibfnamefont{E.}~\bibnamefont{Mátyus}},
  \bibinfo{author}{\bibfnamefont{C.}~\bibnamefont{Fábri}},
  \bibinfo{author}{\bibfnamefont{T.}~\bibnamefont{Szidarovszky}},
  \bibinfo{author}{\bibfnamefont{G.}~\bibnamefont{Czakó}},
  \bibinfo{author}{\bibfnamefont{W.~D.} \bibnamefont{Allen}}, \bibnamefont{and}
  \bibinfo{author}{\bibfnamefont{A.~G.} \bibnamefont{Császár}},
  \bibinfo{journal}{J. Chem. Phys.} \textbf{\bibinfo{volume}{133}},
  \bibinfo{pages}{034113} (\bibinfo{year}{2010}).

\bibitem[{\citenamefont{Hazi and Taylor}(1970)}]{stabilization}
\bibinfo{author}{\bibfnamefont{A.~U.} \bibnamefont{Hazi}} \bibnamefont{and}
  \bibinfo{author}{\bibfnamefont{H.~S.} \bibnamefont{Taylor}},
  \bibinfo{journal}{Phys. Rev. A} \textbf{\bibinfo{volume}{1}},
  \bibinfo{pages}{1109} (\bibinfo{year}{1970}).

\end{thebibliography}

\end{document}